\documentclass[conference]{IEEEtran}
\IEEEoverridecommandlockouts
\usepackage{cite}
\usepackage[section]{placeins}
\usepackage{amsmath,amssymb,amsfonts}
\usepackage{algorithm}
\usepackage[algo2e]{algorithm2e}
\usepackage{algpseudocode}
\usepackage{hyperref}
\usepackage{multirow}
\usepackage{pifont} 
\usepackage{textcomp}
\usepackage[table]{xcolor}
\usepackage{color}
\usepackage{graphicx}
\usepackage{subcaption}
\usepackage{booktabs}
\usepackage{makecell}
\usepackage{adjustbox}
\usepackage{stfloats}
\usepackage[numbers,sort&compress]{natbib}
\def\BibTeX{{\rm B\kern-.05em{\sc i\kern-.025em b}\kern-.08em
    T\kern-.1667em\lower.7ex\hbox{E}\kern-.125emX}}
    
\definecolor{tableheader}{HTML}{E8EDF2}   % Header row: light blue-gray
\definecolor{oursrow}{HTML}{F0F7FF}       % Our method row: very light blue
\definecolor{bestcell}{HTML}{D6EAF8}      % Best cell highlight (optional)

\begin{document}

\title{HiFuzz: Hierarchical Reinforcement Learning for Semantic-Aware and Adaptive CPU Fuzzing}

\author{
\IEEEauthorblockN{
Ya Wang\IEEEauthorrefmark{1},
Hanwei Fan\IEEEauthorrefmark{1},
Zhenguo Liu\IEEEauthorrefmark{2},
Xiaofeng Zhou\IEEEauthorrefmark{1},
Yangdi Lyu\IEEEauthorrefmark{2},
Jiang Xu\IEEEauthorrefmark{2},
and Wei Zhang\IEEEauthorrefmark{1}
}
\IEEEauthorblockA{\IEEEauthorrefmark{1}Hong Kong University of Science and Technology (HKUST), Hong Kong\\
Email: \{ywangmu, hfanah, xzhoubu\}@connect.ust.hk, wei.zhang@ust.hk}
\IEEEauthorblockA{\IEEEauthorrefmark{2}The Hong Kong University of Science and Technology (Guangzhou), China\\
Email: zliu094@connect.hkust-gz.edu.cn, \{yangdilyu, jiang.xu\}@hkust-gz.edu.cn}
}

\maketitle
\thispagestyle{plain}
\pagestyle{plain}

\begin{abstract}
Modern processor verification struggles to reach deep architectural states due to the inefficiencies of traditional mutation-based fuzzing. We propose HiFuzz, a novel hierarchical reinforcement learning framework that replaces mutation with a structured, two-layer generation process: a Program Agent for global layout and a Basic Block Agent for precise instruction filling. To overcome reward sparsity, HiFuzz integrates an adaptive coverage reward mechanism and a semantic-aware basic block encoder providing intrinsic feedback. Extensive evaluations on three real-world RISC-V cores demonstrate that HiFuzz significantly outperforms state-of-the-art fuzzers in coverage and bug detection.
\end{abstract}


\section{Introduction}

As the scaling benefits of Moore's Law diminish, performance gains increasingly rely on complex microarchitectural techniques. The rapid proliferation of open ISAs, notably RISC-V, has amplified this complexity by fostering a diverse ecosystem of custom implementations and extensions. This evolution has triggered a combinatorial explosion in the design state space and widened the verification gap. Functional verification has therefore become the dominant bottleneck in the hardware design flow, accounting for over 70\% of the total research and development cycle. Residual hardware errata, including speculative execution flaws such as Meltdown~\cite{lipp2018meltdown} and Spectre~\cite{kocher2019spectre}, can compromise system security and require costly silicon respins.

Traditional verification methodologies struggle to keep pace with this growth. Constrained-Random Verification~\cite{bergeron2003writing} suffers from low efficiency and high manual overhead, while formal verification~\cite{clarke2018handbook} is computationally intractable for full-scale designs due to state explosion. Hardware fuzzing has therefore emerged as a scalable alternative.

\begin{figure}[t]
    \centering
    \includegraphics[width=0.9\columnwidth]{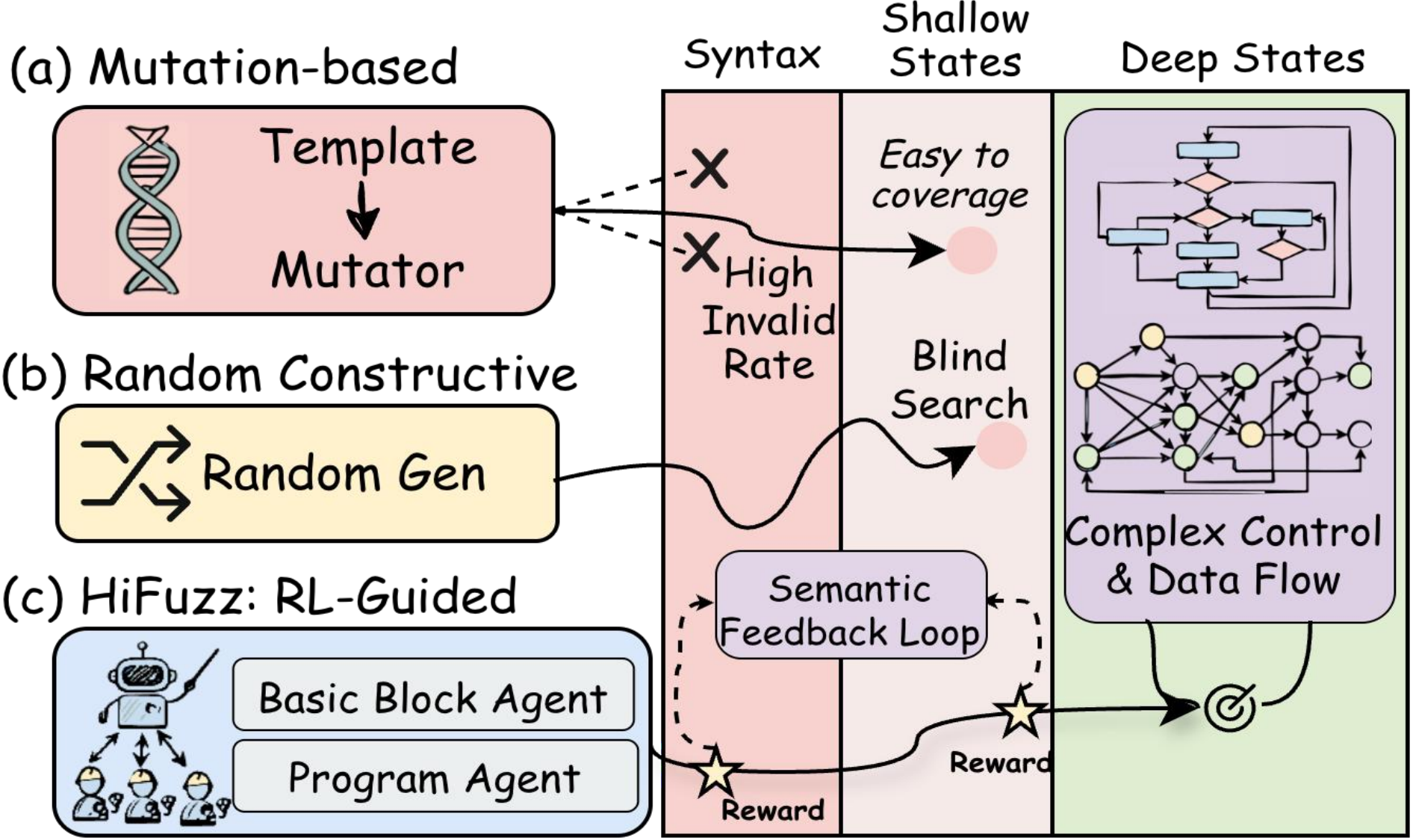}
    \vspace{-2mm}
    \caption{The paradigm shift in hardware fuzzing: mutation-based generation, random constructive generation, and HiFuzz's RL-guided generation with semantic feedback.}
    \label{fig:teaser}
    \vspace{-4mm}
\end{figure}

Fig.~\ref{fig:teaser} illustrates three hardware-fuzzing paradigms. Early fuzzers use mutation-based strategies~\cite{laeufer2018rfuzz,hur2021difuzzrtl}, which generate tests by modifying existing templates. Mutation often disrupts instruction semantics, so many generated programs are discarded before they reach meaningful pipeline stages. Recent work moves toward constructive program generation~\cite{solt2024cascade}, which builds syntactically correct instruction streams from scratch to preserve execution validity. These methods, however, still rely mainly on randomized heuristics. Without learning from prior executions, they explore the state space blindly and struggle to reach deep architectural states.

This gap suggests a role for reinforcement learning, or RL, which can optimize generation strategies from coverage feedback. Integrating RL into constructive generation is harder than applying it to mutation-based fuzzers. In mutation-based approaches, the agent selects a discrete operator for an existing seed, which is a direct control problem. Constructive generation must instead assemble a program instruction by instruction while respecting syntactic and semantic dependencies. This creates a \textit{control-validity dilemma}: the generator must preserve structural integrity while still exposing an expressive action space to the RL agent. Naive RL applications fail because the search space is combinatorial and the validity constraints are rigid.

To use RL for constructive hardware fuzzing, we identify three challenges:
\begin{itemize}
\item \textbf{Complex Action Space and Structural Dependencies:} Unlike seed mutation, constructive generation operates at the level of basic blocks and instruction sequences. Each decision constrains the validity and feasibility of later instructions, so mapping this dependency-laden process into a learnable RL action space without violating program correctness remains an open challenge.
\item \textbf{Sparse Feedback and Delayed Rewards:} Coverage metrics are typically available only after full test program execution, resulting in severe reward sparsity. The agent must make many sequential decisions before receiving any feedback, and the long delay between action and reward complicates credit assignment.
\item \textbf{Reward Bias and Masking Effect:} Optimizing a single aggregated coverage metric often produces a masking effect. The agent can maximize total reward by over-exploiting easy modules while neglecting harder-to-reach components. Without balanced exploration, the policy converges on local optima and misses critical states.
\end{itemize}

To tackle these challenges, we propose \textbf{HiFuzz}. HiFuzz addresses Challenge~1 through a hierarchical decomposition of the generation process, Challenge~2 through a Semantic-Aware Basic Block Encoder that provides intrinsic novelty rewards, and Challenge~3 through an Adaptive Coverage Reward Mechanism based on the Upper Confidence Bound algorithm.

In summary, this paper makes the following contributions:
\begin{itemize}
\item \textbf{A Configurable Test Generation Framework:} We introduce a hierarchical test generation framework that integrates global configurability with basic block-level control. This design systematically constructs valid test cases with complex data and control-flow dependencies, exception handling, and privilege transitions. The generator serves as the foundation for the RL components.
\item \textbf{A Dual-Agent RL Architecture:} We design a hierarchical RL system comprising a Rainbow DQN-based Program Agent for high-level structural configuration and a PPO-based Basic Block Agent for precise instruction generation.
\item \textbf{Semantic-Aware Intrinsic Feedback:} We introduce a Semantic-Aware Basic Block Encoder trained via a novel two-stage pipeline: self-supervised masked language modeling followed by supervised contrastive learning with a custom BB similarity metric. The resulting embedding space quantifies the novelty of generated basic blocks, providing immediate intrinsic rewards without hardware simulation.
\item \textbf{Adaptive Coverage Reward Mechanism:} We develop a UCB-based reward mechanism to mitigate coverage bias and the masking effect. By decomposing coverage at the module level and dynamically weighting each module's contribution, this mechanism ensures balanced exploration across all hardware components.
\item \textbf{Comprehensive Evaluation:} Experiments on three industry-standard RISC-V cores, Rocket, BOOM, and CVA6, show that HiFuzz substantially outperforms state-of-the-art baselines in both coverage convergence and bug detection.
\end{itemize}

\section{Background and Motivation}

\subsection{Hardware Fuzzing}
Hardware fuzzing explores processor states by generating test programs and checking their behavior against a reference model. Early frameworks such as RFUZZ~\cite{laeufer2018rfuzz}, and later DifuzzRTL~\cite{hur2021difuzzrtl} and TheHuzz, use \emph{coverage-guided mutation}. Each test is produced by mutating a seed at the instruction level and then keeping inputs that improve coverage. Blind mutation often breaks the semantic validity of long RISC-V programs. Cascade~\cite{solt2024cascade} reports that DifuzzRTL's stream has a median completion rate of only 1.7\% and a prevalence of 3.0\%, so most instructions never reach the back end. ProcessorFuzz~\cite{canakci2023processorfuzz} augments DifuzzRTL with CSR-transition filtering, but the underlying feedback remains tied to control-register activity. Frequently updated counters such as \texttt{instret} and \texttt{fflags} can dominate this signal, causing the fuzzer to revisit a narrow subset of register states rather than explore broader architectural behavior. Mutation-based loops therefore trade off \emph{validity} against \emph{feedback fidelity} and tend to stall on both.

Generation-based fuzzers reduce this validity problem. Cascade~\cite{solt2024cascade} uses an asymmetric ISA pre-simulation step to synthesize valid instruction streams from scratch while interleaving control- and data-flow construction. Its generation, however, is driven by random heuristics and does not learn from past runs. ChatFuzz~\cite{rostami2024chatfuzz} and GenHuzz~\cite{wu2025genhuzz} inject reinforcement learning into the loop, but they emit a flat token sequence without explicit program structure. This leaves three limitations: the policy is optimized against a specific DUT's coverage counters, recovery from stalled exploration relies on global policy resets, and token-level generation can still emit programs that fail after compilation. A more detailed structural contrast with GenHuzz is deferred to Sec.~\ref{sec:structural-cmp-genhuzz}. These observations motivate two design goals: a reward signal grounded in program semantics rather than a single coverage counter, and an exploration mechanism that redirects effort toward under-verified components without discarding the learned policy.

\subsection{Chisel-Based CPU Design}
Chisel is a Scala-embedded hardware construction language that describes circuits as parameterized generators rather than fixed RTL instances~\cite{bachrach2012chisel}. This high-level representation is widely used in the RISC-V ecosystem. Rocket~\cite{asplos2016rocket} and BOOM~\cite{celio2017boom} are both Chisel-based cores.

The value of a high-level hardware representation is not limited to design productivity. It can also serve verification infrastructure. Laeufer et al.~\cite{laeufer2023simulator} show that coverage metrics such as line, toggle, and finite-state-machine coverage can be implemented as compiler instrumentation passes over FIRRTL, decoupling coverage collection from a specific simulator backend. This observation is important for hardware fuzzing because feedback quality depends on what the instrumentation can expose. HiFuzz uses such high-level coverage structure when it is available on Chisel-generated cores, but the method does not rely on Chisel itself. The CVA6 evaluation uses an independent SystemVerilog design to test whether the generation policy transfers beyond the Rocket Chip ecosystem.

\subsection{Hierarchical Reinforcement Learning}
Hierarchical Reinforcement Learning decomposes a complex decision-making problem into subtasks solved at different temporal or semantic scales~\cite{sutton1999between}. This structure fits constructive program generation, where decisions range from global memory layout to individual instruction selection.
HiFuzz uses a two-level hierarchy. A high-level \textit{Program Agent} determines the global structure and configuration of the test program, while a low-level \textit{Basic Block Agent} generates instruction sequences. This decomposition reduces the action space seen by each agent and enables more efficient learning and exploration than a flat RL policy.

\begin{figure*}[!t]
    \centering
    \includegraphics[width=0.84\textwidth]{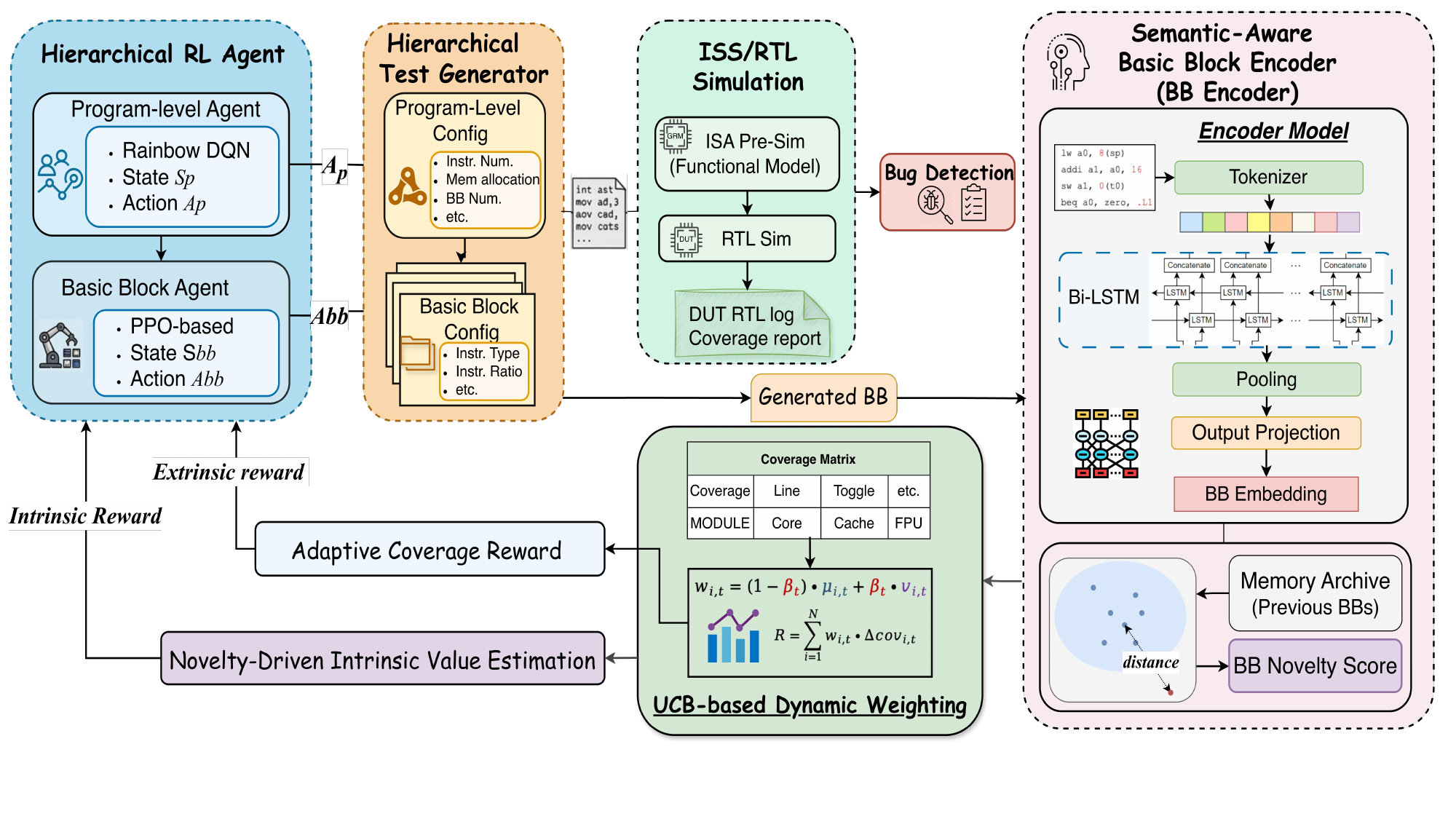}
    \captionsetup{skip=2pt}
    \caption{HiFuzz framework overview. The dual-agent architecture collaboratively generates test programs: (1) Program Agent determines global configuration; (2) Basic Block Agent fills instruction sequences; (3) Semantic-Aware BB Encoder provides intrinsic rewards; (4) Adaptive Coverage Reward Mechanism balances module-level exploration.}
    \label{overview}
\end{figure*}

\section{Framework}

\subsection{Overview}
As shown in Fig.~\ref{overview}, HiFuzz uses a dual-agent architecture together with two feedback mechanisms. The \textbf{Program Agent} acts as the high-level planner, determining global properties such as memory layout, the number of basic blocks (BBs), and the control-flow structure of each test program. The \textbf{Basic Block Agent} functions as the low-level executor, filling each BB with valid instruction sequences according to the configuration handed down by the Program Agent.

Two feedback signals close the loop. The \textbf{Semantic-Aware Basic Block Encoder} (Section~\ref{subsec:bb-encoder}) returns an immediate intrinsic novelty reward after each BB is produced, so the Basic Block Agent receives a dense learning signal without waiting for RTL simulation. The \textbf{Adaptive Coverage Reward Mechanism} (Section~\ref{subsec:acrm}) consumes post-simulation coverage and dynamically rebalances module-level weights so that hard-to-reach components do not get starved by easier ones.

\subsection{Configurable Test Generator}
The foundation of HiFuzz is a highly configurable test generator capable of producing syntactically and semantically valid RISC-V programs. Building upon the principles of constructive generation, our generator introduces a hierarchical configuration mechanism that separates global structural decisions from local instruction details.

\begin{figure}[!t]
    \centering
    \includegraphics[width=0.95\linewidth]{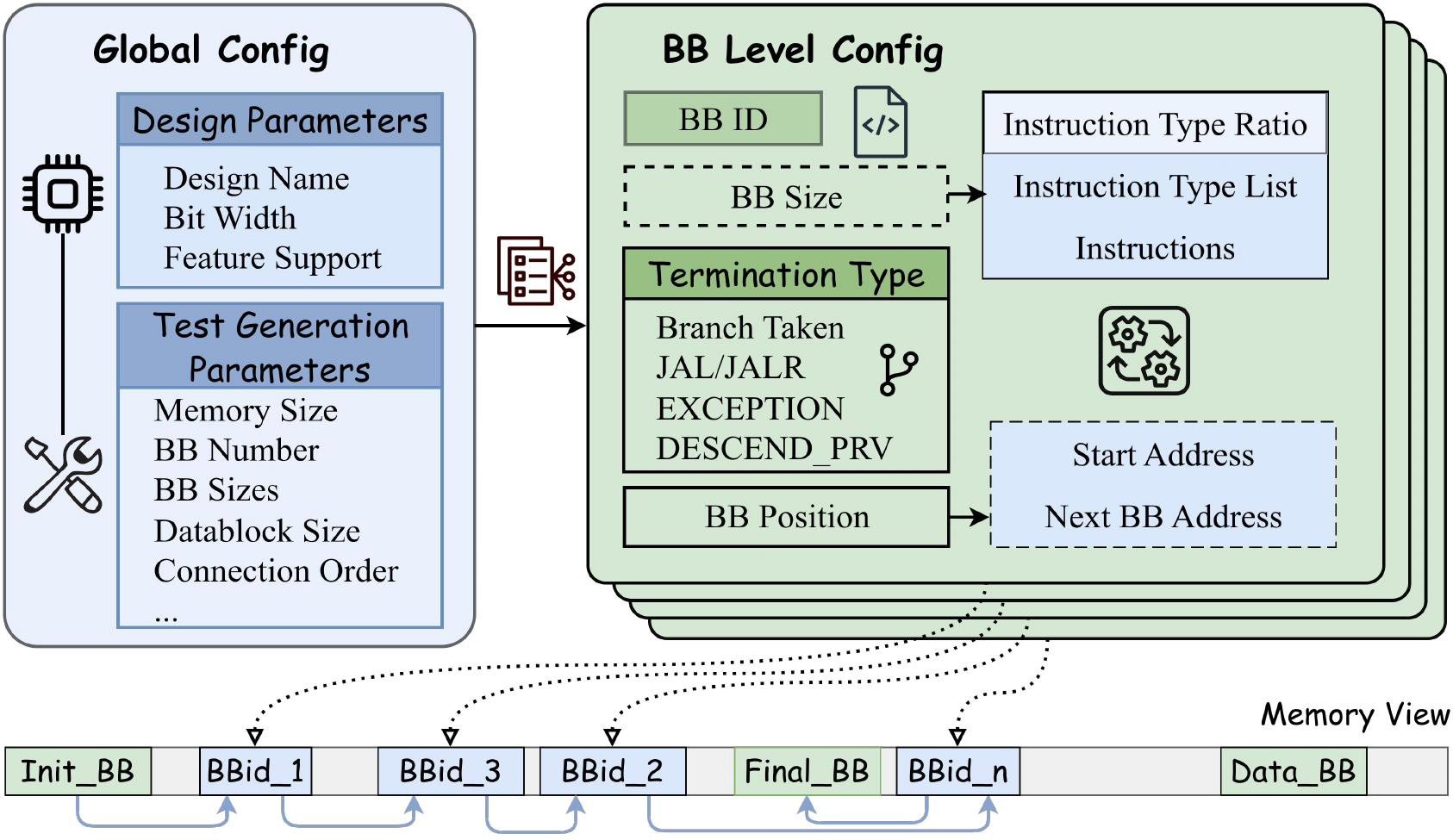}
    \caption{Hierarchical Configuration mechanism}
    \label{fig:config}
\end{figure}

Figure~\ref{fig:config} summarizes the generation workflow. Given a \textit{Global Config}, HiFuzz first fixes the macro structure of the program: total memory budget, BB count, BB-size distribution, and the control-flow template that links the BBs. It then allocates aligned, non-overlapping code and data regions and assigns each BB concrete start, end, and successor addresses consistent with that global layout. For each BB, the corresponding \textit{BB Level Config} specifies the target instruction-category mix (integer, floating-point, memory, CSR, etc.) together with the termination mode (branch, jump, or exception); the constrained generator then instantiates concrete instructions while enforcing operand dependencies, privilege constraints, address legality, and jump-target consistency. The final program is accepted only after whole-program validation confirms that every transfer target stays within allocated regions and every memory access remains in bounds, after which the executable test is sent to the ISA pre-simulator and RTL run-time environment. This hierarchical split between global structure and per-BB content lets HiFuzz explore the structural space of programs systematically while retaining the fine-grained control needed to trigger specific micro-architectural behaviors.

\subsection{HRL Agent Architecture}\label{subsec:hrl-agent}
\begin{figure}[!htbp]
    \centering
    \includegraphics[width=0.95\linewidth]{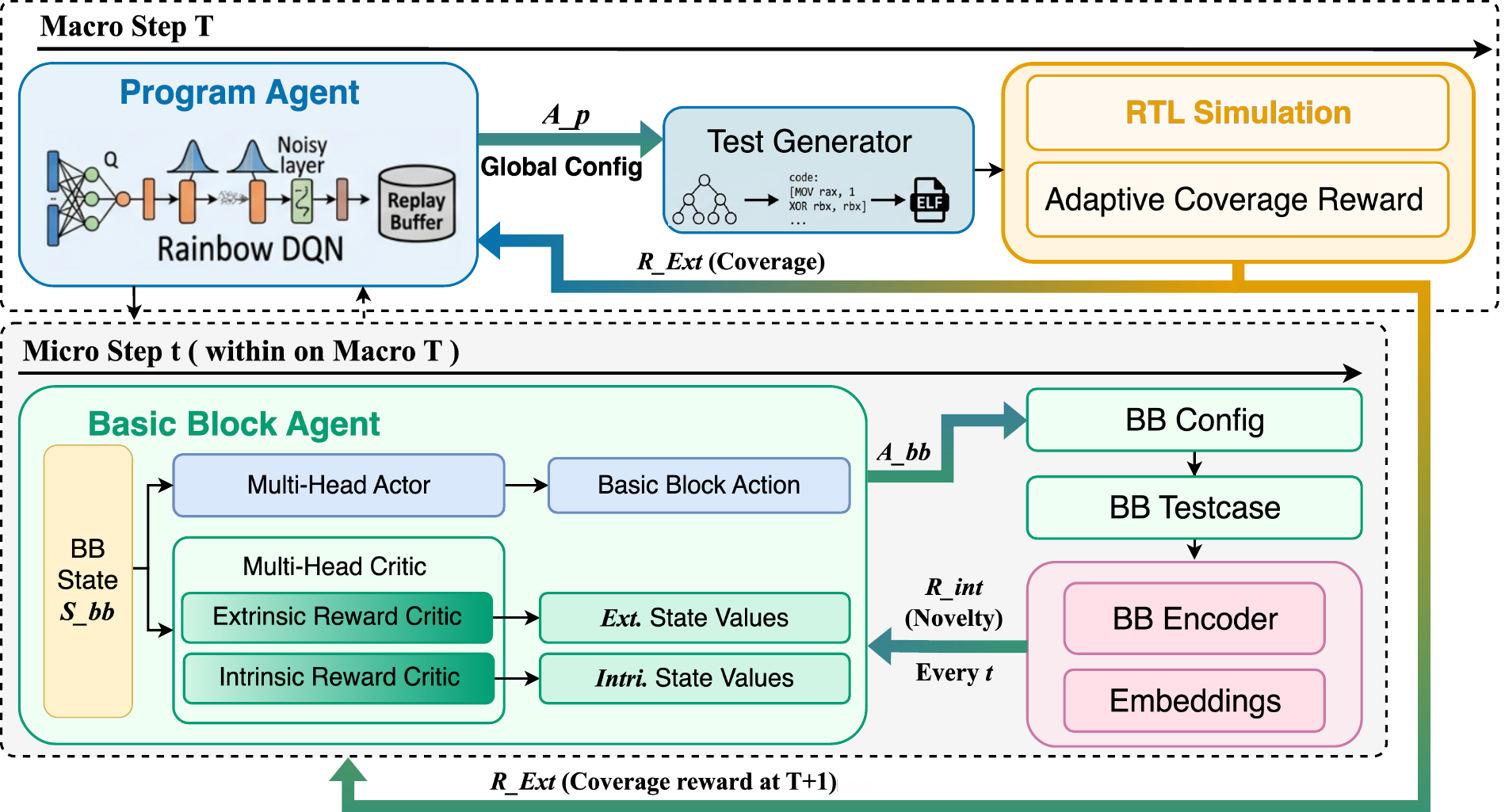}
    \caption{HRL Agent Interaction (Temporal Workflow). The Program Agent selects global config at macro step $T$; the Basic Block Agent fills each BB at micro steps $t=1,\ldots,N$. Intrinsic reward $R_{int}$ flows from BB Encoder to BB Agent; extrinsic reward $R_{ext}$ flows from CPM to Program Agent at $T+1$.}
    \label{fig:hrl_interaction}
\end{figure}

HiFuzz employs a two-level hierarchical RL architecture (Fig.~\ref{fig:hrl_interaction}) that decomposes the complex program-generation task into two decoupled subproblems. The high-level Program Agent determines the global structure of the test program, while the low-level Basic Block Agent decides the instructions inside each BB. This decomposition is motivated by the fact that program-level decisions (memory layout, BB count, control flow) operate on a fundamentally different timescale and abstraction level from instruction-level decisions (operand selection, dependency management, BB termination), so assigning them to a single flat policy would mix signals that benefit from very different learning dynamics.

The two agents interact through a shared environment: the Program Agent's macro action defines the context in which the Basic Block Agent operates, while the Basic Block Agent's micro actions produce the concrete test programs whose coverage outcomes, in turn, drive the learning signals of both agents.

\subsubsection{Program Agent}\label{subsubsec:program-agent}
The Program Agent acts as a \emph{structural configuration scheduler}: at the start of each episode it picks a global layout for the test program, including memory footprint, BB count, per-BB length distribution, and privilege mix, without ever choosing individual instructions. Its input is a compact 11-dimensional summary of recent generation statistics (BB/instruction completion rates, BB-size moments, and training progress), and its output is one of 1{,}375 discrete configurations formed by the cross-product of memory, count, length, and distribution levels. We use Rainbow DQN~\cite{hessel2018rainbow} for this role because its prioritized replay and $n$-step returns are well matched to a discrete action space with delayed, sparse coverage rewards. Exact state fields and action levels are tabulated in Appendix~\ref{sec:impl_appendix}.

\subsubsection{Basic Block Agent}\label{subsubsec:bb-agent}
The Basic Block Agent is trained with Proximal Policy Optimization (PPO)~\cite{schulman2017ppo} under a multi-head Actor-Critic architecture: the Actor emits the BB-level configuration (instruction-type distribution and BB-termination signal), while the Critic carries \emph{separate} value heads for intrinsic and extrinsic returns so that reward streams of different scales and variances are not collapsed into a single target. Letting the agent choose concrete instructions directly would expose it to a combinatorial action space whose size grows with both BB length and ISA breadth, which is impractical for any on-policy RL algorithm. We therefore have the agent output only a compact set of category-level distribution parameters and delegate syntactic and semantic correctness to the Cascade-based constrained generator; this abstraction keeps the policy tractable while still giving it enough control to shape the instruction mix and termination of each BB.

To counter the sparsity of the extrinsic coverage reward, we complement it with an intrinsic novelty signal inspired by Random Network Distillation (RND)~\cite{burda2019rnd}. Unlike vanilla RND, which measures novelty as the prediction error of a fixed \emph{random} target network, our formulation replaces the random projection with the Semantic-Aware Basic Block Encoder (Section~\ref{subsec:bb-encoder}), so novelty is computed on learned, micro-architecture-aware features rather than arbitrary random ones.

\textbf{Reward Design.} The intrinsic reward $R_{\text{int}}$ is obtained by cluster-distance novelty on the frozen BB Encoder embeddings:
\begin{equation}
R_{\text{int}} = \min_{k}\, \bigl(1 - \cos(E(bb),\, C_k)\bigr),
\label{eq:r_int}
\end{equation}
where $E(bb)$ is the L2-normalized embedding of the generated BB and $\{C_k\}$ are online-updated cluster centers. The extrinsic reward $R_{\text{ext}}$ is derived from coverage improvement normalized against a dynamic baseline:
\begin{equation}
R_{\text{ext}} = \frac{\Delta\mathrm{cov} - \mathrm{baseline}}{\mathrm{baseline}}, \quad \mathrm{baseline} = \bar{R}_{\text{buffer}} \cdot \gamma,
\label{eq:r_ext}
\end{equation}
with $\bar{R}_{\text{buffer}}$ the mean of recent rewards and $\gamma\in(0,1)$ a diminishing factor.

\textbf{Loss Functions.} The agent is trained with the PPO clipped-surrogate objective~\cite{schulman2017ppo}; our design contribution is a \emph{dual-advantage} actor loss that keeps the two reward streams separated all the way through learning:
\begin{equation}
\mathcal{L}_{\text{actor}} = \alpha_{\text{int}}\,\hat{A}_{\text{int}} + \alpha_{\text{ext}}\,\hat{A}_{\text{ext}},
\label{eq:l_actor}
\end{equation}
where $\hat{A}_{\text{int}}$ and $\hat{A}_{\text{ext}}$ are Generalized Advantage Estimation (GAE)~\cite{schulman2015high} estimates from the two critic heads and $\alpha_{\text{int}}, \alpha_{\text{ext}}$ weight the two streams. The total objective augments $\mathcal{L}_{\text{actor}}$ with the per-head critic MSE losses and an entropy bonus:
\begin{equation}
\mathcal{L} = \mathcal{L}_{\text{actor}} + \sum_{k\in\{\text{int},\text{ext}\}}\!\text{MSE}\bigl(V_k(s_t), y_k\bigr) - \beta\,H\bigl(\pi(\cdot|s_t)\bigr),
\label{eq:l_total}
\end{equation}
where $y_{\text{int}}$ and $y_{\text{ext}}$ are the discounted returns for each stream and $\beta$ controls exploration.

\subsection{Semantic-Aware Basic Block Encoder}\label{subsec:bb-encoder}
Instruction-level RL suffers from a fundamental feedback problem: meaningful coverage is only available after the \emph{entire} program has been generated and simulated on the DUT, which leaves each intra-BB decision without a local training signal. To break this dependency, we introduce a \textbf{Semantic-Aware Basic Block Encoder} that quantifies the novelty of a generated BB directly in an ISA semantic space, without RTL simulation. The encoder is deliberately designed to be DUT-agnostic, so the same trained model can be deployed across Rocket, BOOM, and CVA6 without retraining.

The encoder has three components: (i) a structured tokenizer that converts RISC-V instructions into semantically tagged token sequences, (ii) a Bi-LSTM encoder backbone that produces fixed-dimensional embeddings, and (iii) a two-stage training pipeline that first learns assembly syntax via self-supervised pre-training and then aligns embeddings with a micro-architecture-aware similarity metric. During fuzzing, the frozen encoder drives an online cluster-distance novelty estimator whose output is the intrinsic reward $R_{\mathrm{int}}$ used by the Basic Block Agent (Eq.~\ref{eq:r_int}).

\subsubsection{Structured RISC-V Tokenizer}
\begin{figure}[ht]
    \centering
    \includegraphics[width=0.95\linewidth]{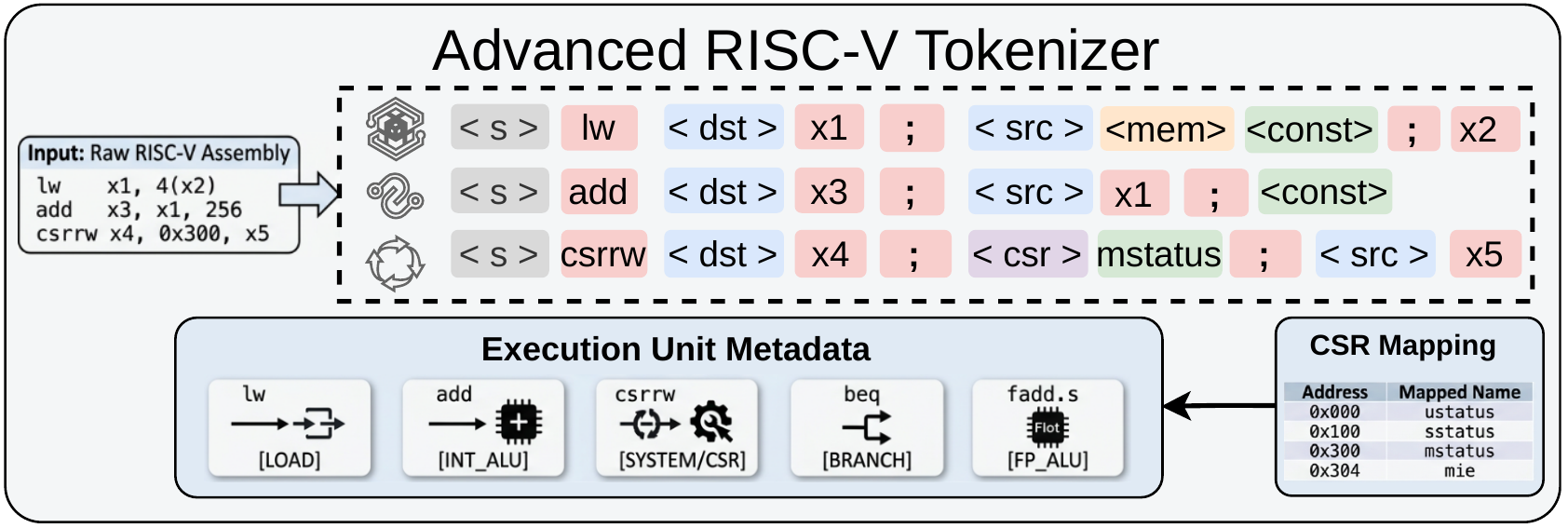}
    \caption{Structured RISC-V tokenizer. Each instruction is decomposed into semantic sub-tokens that expose operand roles, execution-unit metadata, and symbolic CSR names.}
    \label{fig:tokenizer}
\end{figure}

Raw assembly text is a poor input for a learned encoder because treating each instruction as an opaque string discards operand roles, execution-unit metadata, and register structure, which help distinguish micro-architectural behaviors. Our tokenizer (Fig.~\ref{fig:tokenizer}) instead decomposes every instruction into semantic sub-tokens that expose (i) operand roles (destination, source, memory address, CSR, immediate), (ii) the target execution unit annotated per opcode (such as \texttt{INT\_ALU}, \texttt{LOAD}, \texttt{STORE}, \texttt{BRANCH}, \texttt{FPU}), and (iii) symbolic names for control-status registers (such as \texttt{0x300}$\rightarrow$\texttt{mstatus}). Special markers (\texttt{<s>}/\texttt{</s>}, \texttt{<dsts>}/\texttt{<srcs>}, \texttt{<mem>}, \texttt{<csr>}, \texttt{<const>}) delimit these fields.

This design yields two concrete advantages over the token-level scheme used by GenHuzz~\cite{wu2025genhuzz}, which treats each instruction as a whitespace-separated bag of word-level tokens. First, exposing operand roles and execution-unit metadata lets the encoder learn \emph{hardware-aware} relations between instructions---e.g., that two opcodes dispatching to the same functional unit are closer in the embedding space---rather than relying purely on statistical co-occurrence. Second, our vocabulary collapses to only $343$ tokens while still covering 100\% of the RISC-V G ISA, which keeps the embedding table small and stabilizes training of the downstream encoder.

\subsubsection{Encoder Architecture}
We adopt a \textbf{bidirectional LSTM (Bi-LSTM)} as the encoder backbone. The choice is motivated by a structural match between the inductive bias of the model and the object being encoded: a basic block is a short, strictly ordered, single-entry single-exit instruction sequence whose semantics are dominated by local data and control dependencies, whereas attention-based models are built to amortize parameters over long-range context that a BB simply does not contain. Under our two-stage, train-from-scratch regime this mismatch also has a practical cost---Transformer- and linear-attention-based backbones pay for capacity we do not use and are harder to optimize on a moderate-sized corpus without large-scale pretraining, while a Bi-LSTM converges more stably on the same data. To confirm that no strong candidate is being left on the table, we benchmarked two additional families of sequence models---Transformer-based (GPT-2, GPT-2-small) and linear-attention (RWKV-7)---and report the full size/latency comparison in Appendix~\ref{sec:encoder_comparison}.

\subsubsection{Two-Stage Training Process}
\begin{figure}[ht]
    \centering
    \includegraphics[width=0.95\linewidth]{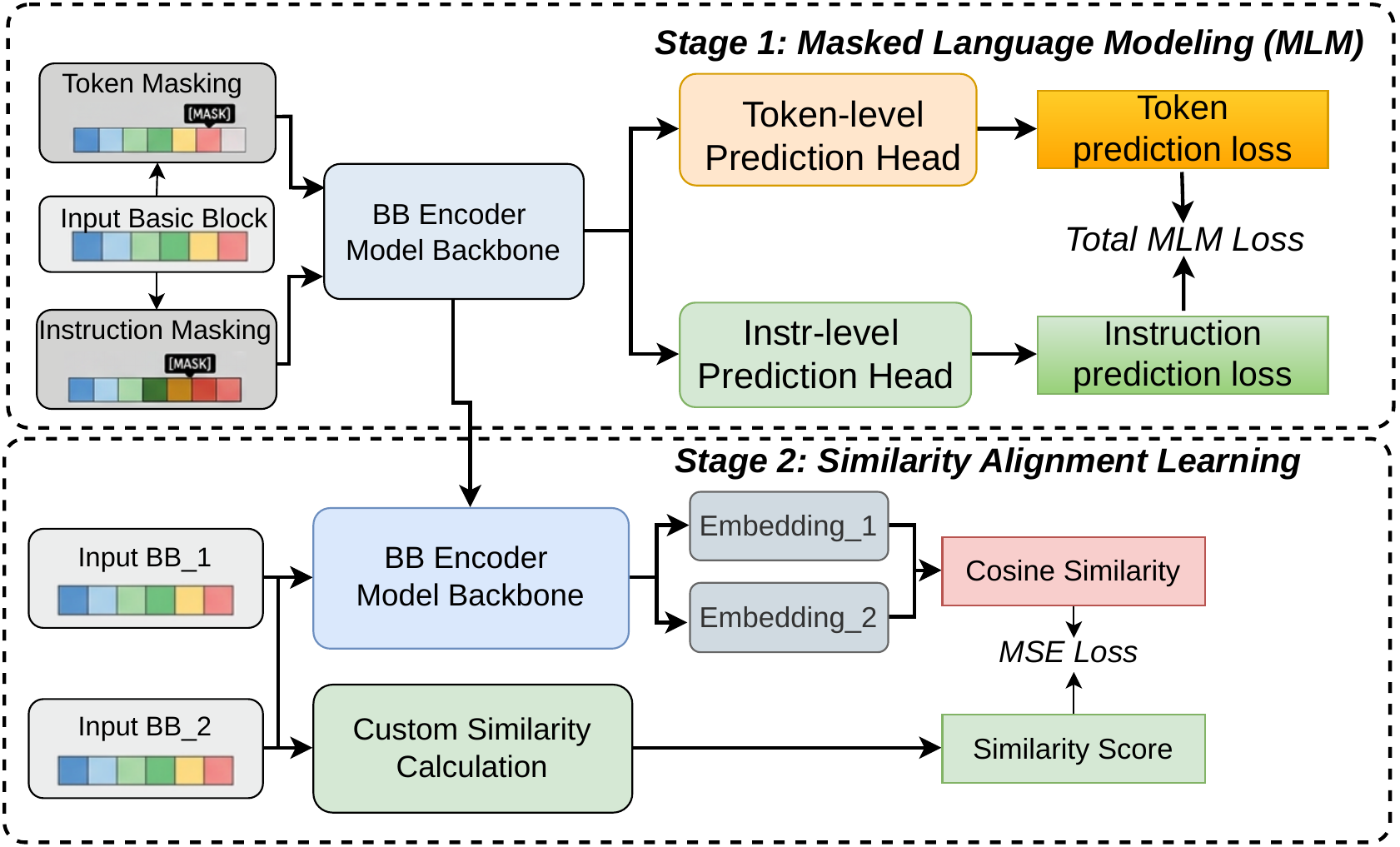}
    \caption{Two-stage training pipeline for the BB Encoder.}
    \label{fig:bb_encoder}
\end{figure}

We employ a curriculum learning approach, moving from syntax and grammar learning to similarity discrimination. Fig.~\ref{fig:bb_encoder} illustrates the complete training pipeline.

\textbf{Stage 1: Self-Supervised Syntax Learning (MLM).}
Stage~1 teaches the encoder the syntax and structural patterns of RISC-V assembly from $50$K$+$ unlabeled samples, without manual annotations. We use a dual-level masking strategy: \textit{token-level masking} randomly masks 15\% of individual tokens (operands, registers) so the model learns local usage patterns, while \textit{instruction-level masking} masks entire instructions so the model learns sequential dependencies across instructions. This self-supervised objective lets the encoder internalize RISC-V semantics before any task-specific fine-tuning.

\textbf{Stage 2: Supervised Similarity Learning.}
Stage~2 aligns the embedding space with a micro-architecture-aware similarity metric. In software testing, functional equivalence (identical input/output mappings) provides natural similarity labels: for example, different compiler optimizations of the same source program are still ``similar'' despite differing instruction sequences. In hardware fuzzing, however, functionally equivalent programs may stress entirely different internal CPU components; for example, unoptimized memory accesses and optimized register-only operations can exercise different execution units and dependency hazards.

To bridge this gap, we define a custom structural similarity metric, \textit{BB-Sim} (Algorithm~\ref{alg:bb_sim}), that draws on the weighted sequence-alignment intuition used in binary-difference work such as BinSequence~\cite{huang2017binsequence}, but extends it with hardware-specific features: instruction-type overlap, operand register reuse patterns, and data-dependency structure. This metric provides supervision for training the encoder to distinguish instruction sequences that stress different hardware components, even when they produce identical functional behavior.

\begin{algorithm}[ht]
    \caption{Basic Block Similarity Calculation (BB-Sim)}
    \label{alg:bb_sim}
    \footnotesize
    \begin{minipage}{\linewidth}
    \raggedright
    \textbf{Input:} Basic blocks $A=(a_1,\ldots,a_n)$ and $B=(b_1,\ldots,b_m)$\\
    \textbf{Output:} Similarity score $S$\\
    \textit{// 1. Initialize DP table}\\
    \hspace*{1em}Initialize $DP[0{:}n][0{:}m] \leftarrow 0$\\
    \textit{// 2. Compute weighted sequence score}\\
    \hspace*{1em}\textbf{for} $i \leftarrow 1$ \textbf{to} $n$ \textbf{do}\\
    \hspace*{2em}\textbf{for} $j \leftarrow 1$ \textbf{to} $m$ \textbf{do}\\
    \hspace*{3em}$s \leftarrow \textsc{MatchScore}(a_i,b_j)$\\
    \hspace*{3em}$s_{\mathrm{diag}} \leftarrow DP[i-1,j-1] + s$\\
    \hspace*{3em}$DP[i,j] \leftarrow \max\{DP[i-1,j],\,DP[i,j-1],\,s_{\mathrm{diag}}\}$\\
    \hspace*{2em}\textbf{end}\\
    \hspace*{1em}\textbf{end}\\
    \hspace*{1em}$S_{\mathrm{seq}} \leftarrow DP[n,m]$\\
    \textit{// 3. Add dependency bonus}\\
    \hspace*{1em}$P \leftarrow \textsc{BacktrackPairs}(DP,A,B)$\\
    \hspace*{1em}$S_{\mathrm{dep}} \leftarrow 0$\\
    \hspace*{1em}\textbf{foreach} $(i,j) \in P$ \textbf{do}\\
    \hspace*{2em}$S_{\mathrm{dep}} \leftarrow S_{\mathrm{dep}} + \textsc{DepBonus}(a_i,b_j,W)$\\
    \hspace*{1em}\textbf{end}\\
    \textit{// 4. Normalize}\\
    \hspace*{1em}\textbf{return} $(S_{\mathrm{seq}} + S_{\mathrm{dep}})/\max(n,m)$
    \end{minipage}
\end{algorithm}

Algorithm~\ref{alg:bb_sim} keeps the full computation path while abbreviating only the low-level scoring rules. Specifically, $\textsc{MatchScore}(a_i,b_j)$ first assigns $B_{\mathrm{exact}}$ to an exact opcode match, or a discounted $\alpha \cdot B_{\mathrm{exact}}$ when the two instructions target the same execution unit; it then adds $B_{\mathrm{sameop}}$ for each operand position whose type matches. The helper $\textsc{DepBonus}(a_i,b_j,W)$ contributes $\beta_{\mathrm{bonus}}$ when the aligned instruction pair exhibits consistent dependency behavior within window $W$. This preserves the weighted sequence-alignment intuition while keeping the pseudocode compact enough for the main text.

In Stage 2, we employ contrastive learning where the cosine similarity of the generated embeddings approximates this custom similarity score $S$. The encoder is trained to minimize the MSE loss: $\mathcal{L} = (\cos(E(BB_1), E(BB_2)) - S)^2$. By forcing the network to predict this micro-architecturally grounded metric, the resulting semantic embedding space inherently reflects hardware stress patterns. Consequently, vector closeness in this space directly corresponds to structural and hardware-utilization similarity, providing the intrinsic reward signal for the RL agent.

\subsection{Adaptive Coverage Reward Mechanism}\label{subsec:acrm}
When the agent is optimized against a single global coverage metric, easily covered modules tend to dominate the reward signal, mask the starvation of harder-to-reach components, and drive the policy toward local optima. HiFuzz addresses this with a module-aware Adaptive Coverage Reward Mechanism (ACRM) based on the Upper Confidence Bound algorithm~\cite{auer2002finite}: coverage is decomposed at the module level and each module's contribution is reweighted dynamically so that exploration stays balanced across the entire design.

\subsubsection{Module Breakdown Strategy}
Instead of optimizing a single global coverage number, we slice the DUT into $M$ independent sub-modules and aggregate coverage metrics (Line, Toggle, MUX, Control Register) at the granularity of each module (e.g., core.div, dcache, fpuOpt, ptw). Let $\text{Cov}_i^{(t)}$ denote the coverage of module $i \in \{1, \dots, M\}$ at step $t$. This allows us to track the verification status of each component independently and ensure balanced exploration.

\subsubsection{Dynamic Weighting with UCB}
We formulate the module selection as a Multi-Armed Bandit problem. The weight $w_i^{(t)}$ for module $i$ at step $t$ is calculated dynamically using a UCB-based approach:
\begin{equation}
    \tilde{w}_i^{(t)} = \mu_i^{(t)} + \beta^{(t)} \cdot v_i^{(t)}
\end{equation}
\begin{equation}
    w_i^{(t)} = \frac{\exp(\tilde{w}_i^{(t)} / \tau)}{\sum_{j=1}^M \exp(\tilde{w}_j^{(t)} / \tau)}
\end{equation}
where:
\begin{itemize}
    \item $\mu_i^{(t)}$ is the historical mean coverage gain for module $i$ (Exploitation).
    \item $v_i^{(t)} = 1 - \frac{\text{current\_cov}_i^{(t)}}{\text{max\_cov}_i}$ represents the saturation level of the module (Exploration).
    \item $\beta^{(t)}$ is a dynamic parameter that grows as training progresses, shifting focus from exploiting high-yield modules to exploring under-verified ones.
    \item $\tau$ is a temperature parameter for softmax normalization.
\end{itemize}
The final extrinsic reward $R_{ext}^{(t)}$ passed to the Program Agent is a weighted sum of coverage gains across all modules: 
\begin{equation}
    R_{ext}^{(t)} = \sum_{i=1}^M w_i^{(t)} \cdot \Delta \text{Cov}_i^{(t)}
\end{equation}
This mechanism ensures that the fuzzer continuously shifts its focus to under-covered modules, preventing local optima.

\section{Implementation Details}\label{sec:impl-details}
HiFuzz is implemented in Python on top of Cocotb, Verilator~\cite{verilator2024}, and Spike, and is evaluated on Rocket~\cite{asplos2016rocket}, BOOM~\cite{celio2017boom}, and CVA6~\cite{zaruba2019cva6}. The Program Agent is trained with Rainbow DQN~\cite{hessel2018rainbow} over an 11-dimensional generation-statistics state and a $1{,}375$-way discrete configuration action; the Basic Block Agent is trained with PPO~\cite{schulman2017ppo} on top of a multi-head Actor--Critic, using intrinsic/extrinsic reward coefficients $\alpha_{\text{int}}{=}1$, $\alpha_{\text{ext}}{=}2$ and extrinsic-baseline diminishing factor $\gamma{=}0.75$. Both agents are narrow MLPs trained with learning rates $1\mathrm{e}{-4}$ and $1\mathrm{e}{-3}$, respectively, on an Intel Xeon Gold 6246R server with one NVIDIA A100 GPU. Full state/action schemas, per-layer network sizes, remaining PPO hyperparameters, coverage-instrumentation primitives, and the YAML-driven experiment configuration are listed in Appendix~\ref{sec:impl_appendix}.

\section{Comprehensive Evaluation}
This section evaluates the performance of HiFuzz on real-world RISC-V processors. We first detail the evaluation setup, followed by experiments assessing coverage efficiency, ablation studies, program generation quality, BB Encoder effectiveness, and bug detection capability.

\subsection{Experimental Setup}
All fuzzing sessions run for 24 hours per DUT on the hardware and with the hyperparameters listed in Section~\ref{sec:impl-details}.

\subsubsection{DUT Processors}
We evaluate three open-source RISC-V cores: \textbf{Rocket Core}~\cite{asplos2016rocket}, an in-order Chisel core; \textbf{BOOM Core}~\cite{celio2017boom}, an out-of-order Chisel core; and \textbf{CVA6}~\cite{zaruba2019cva6}, a six-stage SystemVerilog core with a separate fpnew FPU. Our experiments target the enabled RISC-V G-subset instructions used by these configurations.

CVA6 is independent from the Rocket Chip ecosystem in both HDL and microarchitecture, so it serves as a cross-DUT check rather than another Rocket-family configuration. The reference model for all three cores is Spike.

\subsubsection{Baselines and Metrics}
We compare HiFuzz against three state-of-the-art CPU fuzzers: \textbf{DifuzzRTL}~\cite{hur2021difuzzrtl}, \textbf{ProcessorFuzz}~\cite{canakci2023processorfuzz}, and \textbf{Cascade}~\cite{solt2024cascade}. The coverage metrics used in our evaluation are summarized in Table~\ref{tab:coverage_metrics}. Ablation variants disable the Adaptive Coverage Reward Mechanism (ACRM) or the BB Encoder to isolate each component.

\begin{table}[!htbp]
\centering
\caption{Coverage metrics used in the evaluation.}
\label{tab:coverage_metrics}
\setlength{\tabcolsep}{3.5pt}
\renewcommand{\arraystretch}{1.15}
\footnotesize
\begin{tabular}{lp{0.64\columnwidth}}
\toprule
\rowcolor{tableheader}
\textbf{Metric} & \textbf{Meaning} \\
\midrule
Line & Executed RTL statements. \\
Toggle & Signal value transitions ($0\!\leftrightarrow\!1$). \\
MUX~\cite{laeufer2018rfuzz} & Multiplexer-select toggles. \\
Control Register~\cite{hur2021difuzzrtl} & Exploration of control-register states. \\
\bottomrule
\end{tabular}
\end{table}

These metrics should be interpreted as complementary signals rather than interchangeable scores. Line and Toggle coverage indicate broad RTL activity and are useful sanity checks, but they can saturate even when architecturally interesting states remain unexplored. MUX and Control Register coverage are closer to processor-verification intent because they expose exercised control paths, CSR states, and module-level selection behavior. Consequently, a modest Line/Toggle gain can still be meaningful when accompanied by larger Control Register, MUX, or bug-detection gains; the latter indicate that generated programs are reaching verification-relevant corner cases instead of merely executing more statements.

Unless otherwise stated, each coverage curve reports one full 24-hour end-to-end campaign under a fixed wall-clock budget. We use identical per-fuzzer CPU-core budgets and include generation, RL inference/update, RTL simulation, and coverage collection in the reported time. The \textit{Gen.\ Time} column in Table~\ref{tab:performance_comparison} isolates the time spent outside RTL simulation and coverage collection; the remaining budget is dominated by simulation and coverage processing. The BB Encoder is trained offline once in our workflow and reused across DUTs, so its training cost is not charged to the 24-hour fuzzing budget.

\subsection{Coverage Efficiency}
\subsubsection{Control Register Coverage}

\begin{figure}[!htbp]
    \centering
    \includegraphics[width=\columnwidth]{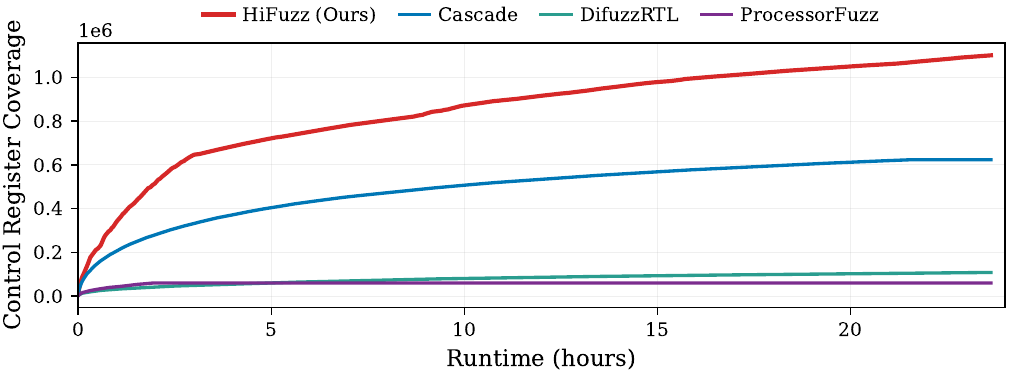}
    \caption{24-hour Control Register Coverage growth on Rocket Core: HiFuzz vs. Cascade, DifuzzRTL, and ProcessorFuzz.}
    \label{fig:ctrlreg_curve}
\end{figure}

We use \textbf{Control Register Coverage} as the primary efficiency metric on Rocket because all three baselines and HiFuzz are evaluated on the same control-register coverage objective. This keeps the comparison focused on generation quality rather than on differences in the reported metric.

\begin{table}[!htbp]
\centering
\caption{24-hour Control Register Coverage on Rocket Core.}
\label{tab:performance_comparison}
\setlength{\tabcolsep}{3.5pt}
\renewcommand{\arraystretch}{1.15}
\footnotesize
\begin{tabular}{l r r r r}
\toprule
\cellcolor{tableheader}\textbf{Method}
  & \cellcolor{tableheader}\textbf{Cov.}
  & \cellcolor{tableheader}\textbf{Tests}
  & \cellcolor{tableheader}\textbf{Gen.\ Time}
  & \cellcolor{tableheader}\textbf{Cov.\ /Test} \\
\midrule
DifuzzRTL~\cite{hur2021difuzzrtl}       & 133,971   & 17,025  & 1.98h (8.3\%)   & 7.87  \\
ProcessorFuzz~\cite{canakci2023processorfuzz} & 103,941   & 26,633  & 12.60h (52.5\%) & 3.90  \\
Cascade~\cite{solt2024cascade}           & 740,329   & 21,439  & 1.06h (4.4\%)   & 34.53 \\
\midrule
DQN+PPO                          & \underline{777,916} & 14,885 & 3.38h (14.1\%) & 52.26 \\
\cellcolor{oursrow}\textbf{HiFuzz (Ours)}
  & \cellcolor{oursrow}\textbf{1,102,343}
  & \cellcolor{oursrow}\textbf{9,454}
  & \cellcolor{oursrow}\textbf{7.32h} (30.5\%)
  & \cellcolor{oursrow}\textbf{116.60} \\
\bottomrule
\end{tabular}
\end{table}

Fig.~\ref{fig:ctrlreg_curve} shows that the mutation-based baselines plateau early on Rocket, while Cascade improves faster at first but flattens later. HiFuzz reaches higher final coverage under the same 24-hour budget. Table~\ref{tab:performance_comparison} shows the same trend in aggregate: compared with Cascade, HiFuzz improves Control Register Coverage by \textbf{48.9\%} and raises coverage per test by \textbf{3.3$\times$}. The gain comes from more effective generated programs, not from a larger test volume.

\subsubsection{Cross-DUT Coverage Efficiency on CVA6}
\begin{figure}[!htbp]
    \centering
    \includegraphics[width=0.95\columnwidth]{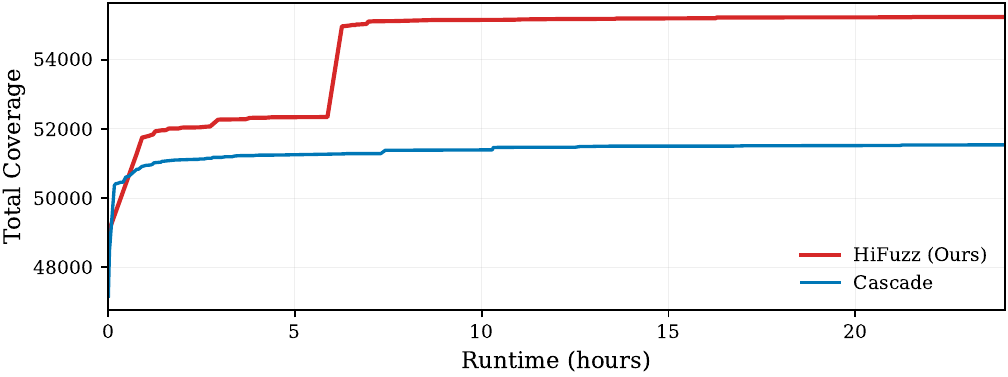}
    \caption{24-hour Total Coverage growth on the SystemVerilog-based CVA6 core: HiFuzz (Ours) vs.\ Cascade.}
    \label{fig:cva6_cov}
\end{figure}

We also test HiFuzz on \textbf{CVA6}, a SystemVerilog DUT outside the Rocket-family implementation stack. Since CVA6 is not written in Chisel, it cannot use the same instrumentation path that we use to collect Control Register Coverage on Rocket. We therefore compare 24-hour \textbf{Total Coverage}, computed as the sum of Verilator Line and Toggle coverage, against Cascade in Fig.~\ref{fig:cva6_cov}. Under the same time budget, HiFuzz reaches 55,241 coverage points, a \textbf{7.2\%} gain over Cascade's 51,541, supporting the generality of the method beyond Rocket-specific RTL structure and instrumentation.

\subsubsection{Module-Level Coverage Breakdown}
\begin{figure}[!htbp]
    \centering
    \includegraphics[width=\columnwidth]{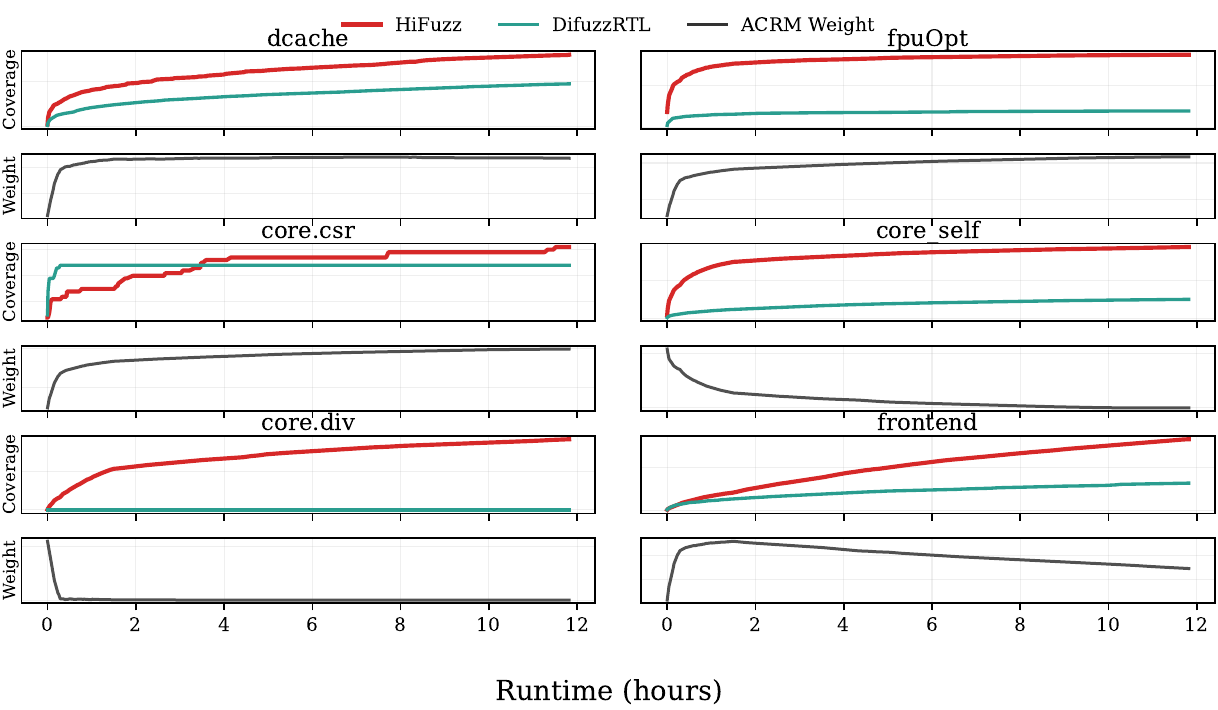}
    \caption{Per-module Control-Register Coverage growth (top) and ACRM weight dynamics (bottom) on six Rocket modules. HiFuzz gains coverage on hard-to-reach modules such as \texttt{fpuOpt}, while ACRM lowers the weight of easier modules after saturation.}
    \label{fig:acrm_combined}
\end{figure}

We next inspect module-level coverage to understand how ACRM changes the search behavior. Fig.~\ref{fig:acrm_combined} reports per-module coverage together with the learned module weights.

The largest gains appear in hard-to-reach modules such as the FPU. As easier modules saturate, the UCB term lowers their reward weight and shifts attention toward under-covered modules instead of letting them be hidden by aggregate coverage. This behavior reduces the masking effect that appears when a single global coverage score is optimized directly.

\subsubsection{MUX, Line, and Toggle Coverage}

\begin{figure}[!htbp]
    \centering
    \includegraphics[width=\columnwidth]{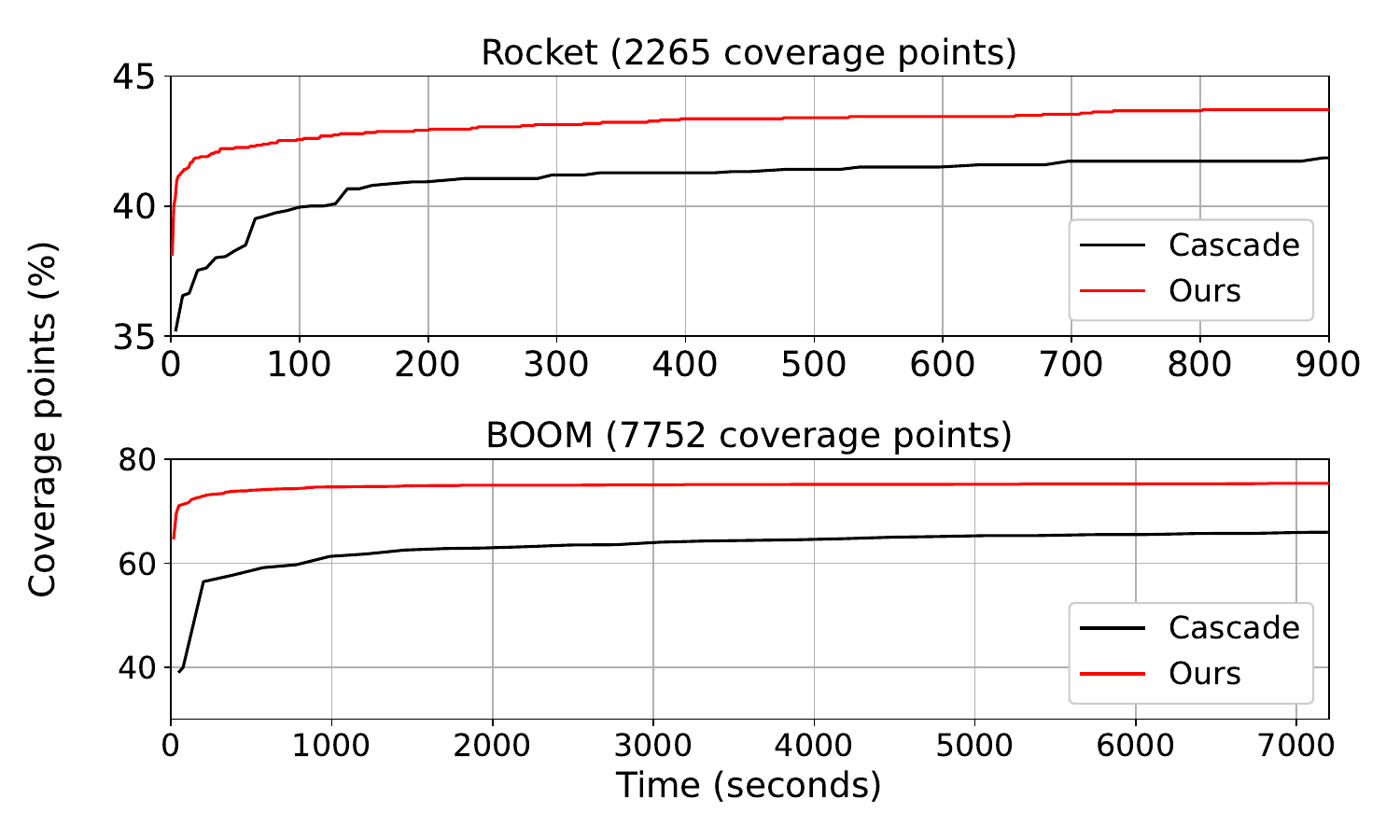}
    \caption{MUX coverage on Rocket and BOOM. HiFuzz's advantage is small on Rocket but widens substantially on the more complex out-of-order BOOM core.}
    \label{fig:muxcoverage}
\end{figure}

\begin{table}[!htbp]
\centering
\caption{Verilator-collected Line and Toggle coverage on Rocket (\%). \textbf{Bold}: best. \underline{Underline}: best baseline.}
\label{tab:avg_coverage_runtime}
\setlength{\tabcolsep}{3.5pt}
\renewcommand{\arraystretch}{1.15}
\footnotesize
\begin{tabular}{lcccc}
\toprule
\rowcolor{tableheader}
\textbf{Metric} & \textbf{DifuzzRTL} & \textbf{ProcFuzz} & \textbf{Cascade} & \textbf{HiFuzz} \\
\midrule
Line   & 72.78 & 72.49 & \underline{73.52} & \textbf{75.98} \\
Toggle & \underline{65.54} & 65.59 & 63.29 & \textbf{67.58} \\
\bottomrule
\end{tabular}
\end{table}

Fig.~\ref{fig:muxcoverage} shows a small MUX-coverage gap on Rocket and a wider gap on BOOM. Table~\ref{tab:avg_coverage_runtime} reports Line and Toggle coverage measured after generation on the same Rocket test cases. These metrics are not used as HiFuzz's training objective, and they tend to saturate quickly on mid-complexity cores~\cite{laeufer2018rfuzz}; the modest gains are therefore expected. The more useful signal is that the advantage grows on the more complex out-of-order DUT.

\subsubsection{Structural Comparison with GenHuzz}\label{sec:structural-cmp-genhuzz}
GenHuzz~\cite{wu2025genhuzz} targets similar RISC-V cores, but its reported coverage uses a different simulation back-end (Synopsys VCS) and an unreleased 12M-parameter model. We therefore avoid a direct coverage-to-coverage comparison in our Verilator-based flow.

The structural contrast is still informative. GenHuzz emits a flat token stream bounded by a preset token budget, so inter-instruction structure must be learned implicitly from next-token statistics on a 5M-instruction random corpus. HiFuzz assigns program length and BB count to the Program Agent and leaves intra-BB instruction choice to the Basic Block Agent. In the program-quality analysis of Fig.~\ref{fig:program_quality}, this design yields an average dependency-chain length of 18.5 and a prevalence of 96.7\%. These are not raw coverage results, but they help explain why HiFuzz can generate denser and more structured tests.

\subsection{Ablation Study}

\begin{figure}[!htbp]
    \centering
    \includegraphics[width=\columnwidth]{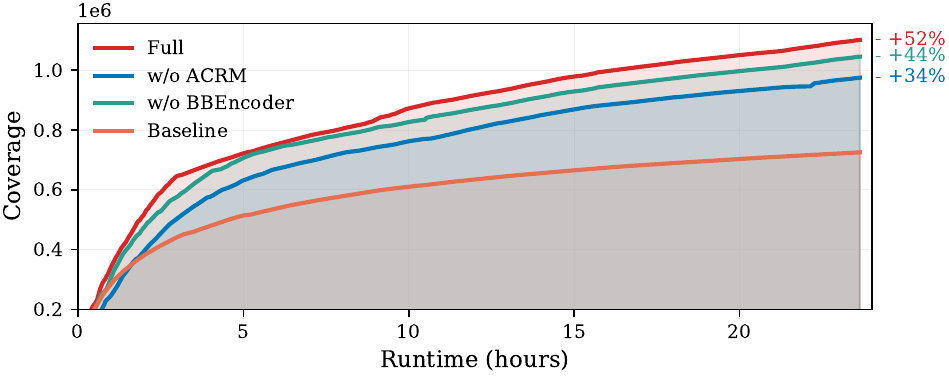}
    \caption{Ablation on Rocket: 24-hour Control Register Coverage growth curves. The BB Encoder contributes $+34.45\%$, ACRM contributes $+44.15\%$, and the full HiFuzz system achieves $+51.87\%$ over the DQN+PPO baseline.}
    \label{fig:ablation}
\end{figure}

We isolate the BB Encoder and ACRM on Rocket. All variants are compared against a DQN+PPO baseline that keeps the dual-agent structure but removes both optimizations.

Fig.~\ref{fig:ablation} shows that the dual-agent baseline alone gives only a small gain over Cascade. Adding the BB Encoder supplies dense intrinsic feedback for instruction-level decisions, while ACRM redirects extrinsic reward toward under-covered modules. The full system achieves a \textbf{51.87\%} improvement over the DQN+PPO baseline, which indicates that the two mechanisms address different parts of the reward problem.

This ablation also clarifies what the experiment does and does not isolate. The DQN+PPO baseline keeps the Program Agent/Basic Block Agent split but removes both semantic intrinsic feedback and adaptive module weighting; therefore, Fig.~\ref{fig:ablation} primarily measures the incremental value of the two reward-shaping mechanisms within the same hierarchical generator. A true single-agent instruction-level RL baseline would require a different action interface and validity mechanism, making it less controlled as a component ablation. We therefore use the DQN+PPO variant to isolate the reward mechanisms, and use the program-quality analysis in Appendix~\ref{sec:prog_quality_appendix} as supporting evidence that the hierarchical generator produces higher-prevalence programs with longer dependency chains than the non-learning and randomly constructive baselines.

\subsection{BB Encoder Performance}
We evaluate the Semantic-Aware Basic Block Encoder (Section~\ref{subsec:bb-encoder}), instantiated as a Bi-LSTM with embedding dimension $96$ and trained on $20{,}000$ Cascade-generated programs. Appendix~\ref{sec:encoder_comparison} compares it with Transformer (GPT-2-style) and Linear-Attention (RWKV) alternatives.

\begin{figure}[!htbp]
    \centering
    \includegraphics[width=\columnwidth]{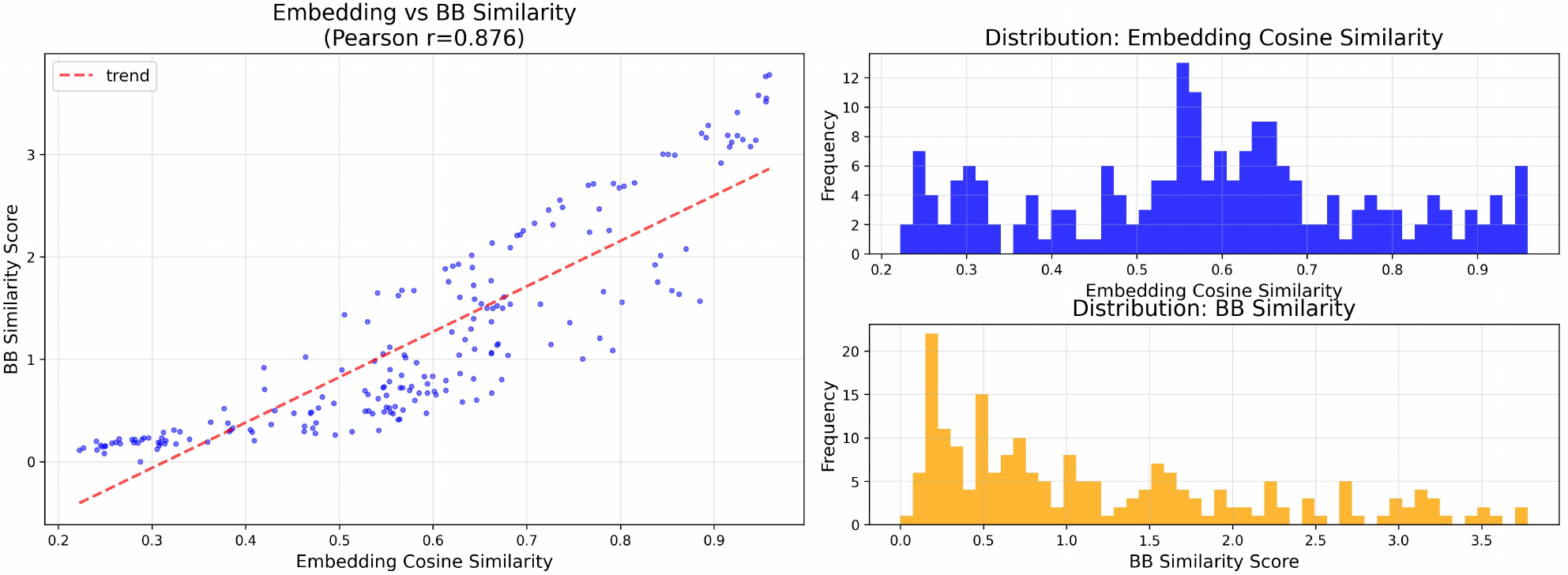}
    \caption{Correlation between embedding cosine similarity and BB similarity score. Pearson $r = 0.876$.}
    \label{fig:bbencoder_corr}
\end{figure}

We measure whether embedding cosine similarity tracks the BB similarity score from Algorithm~\ref{alg:bb_sim}. Fig.~\ref{fig:bbencoder_corr} reports a Pearson correlation of \textbf{0.876}, which supports using the encoder as intrinsic feedback for the BB Agent.

\begin{figure}[!htbp]
    \centering
    \includegraphics[width=\columnwidth]{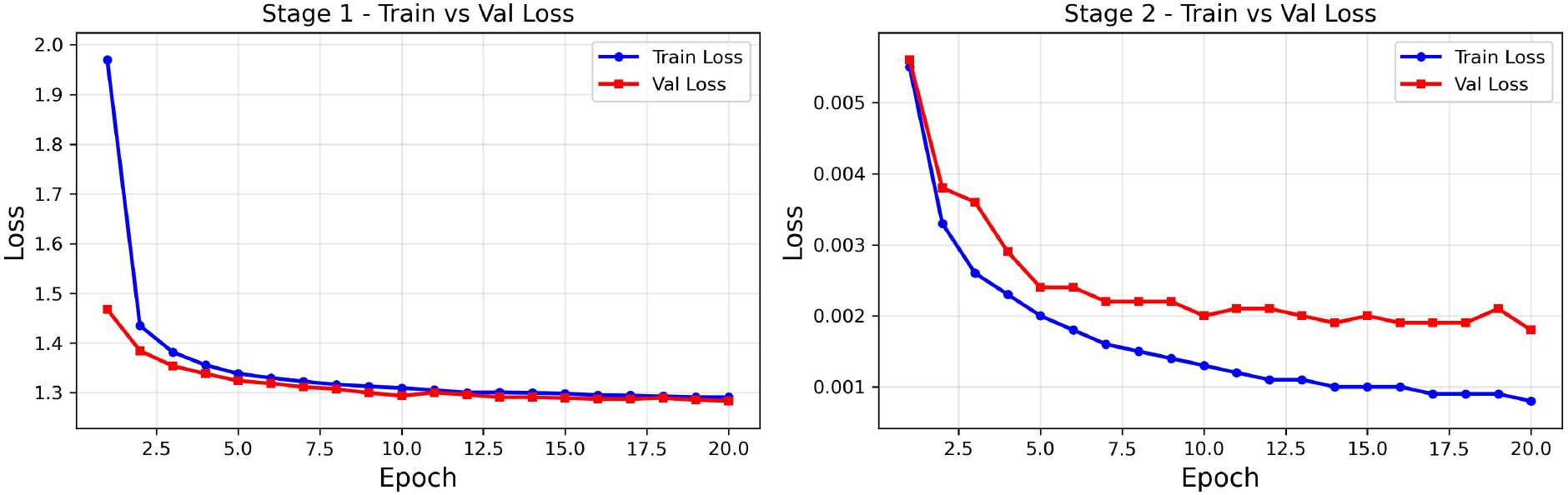}
    \caption{BB Encoder two-stage training loss. Stage~1: Masked Language Modeling. Stage~2: Similarity learning with weighted LCS metric.}
    \label{fig:bbencoder_loss}
\end{figure}

Fig.~\ref{fig:bbencoder_loss} further shows that both stages of the encoder curriculum converge smoothly, including masked language modeling in Stage~1 and similarity learning in Stage~2.

\subsection{Bug Discovery}

\begin{table}[!htbp]
\centering
\caption{Bug detection on the Encarsia benchmark~\cite{bolcskei2025encarsia}: \#detected. Each DUT has 15 Mix-ups (``M'') and 15 Broken Conditionals (``C''); every fuzzer runs 24\,h per bug. \textbf{Bold}: best. \underline{Underline}: best non-HiFuzz baseline (unique).}
\label{tab:bug-detection}
\setlength{\tabcolsep}{2.5pt}
\renewcommand{\arraystretch}{1.15}
\footnotesize
\resizebox{\columnwidth}{!}{%
\begin{tabular}{l ccc ccc c}
\toprule
\rowcolor{tableheader}
 & \multicolumn{3}{c}{\textbf{Rocket}}
 & \multicolumn{3}{c}{\textbf{BOOM}}
 & \textbf{All} \\
\cmidrule(lr){2-4} \cmidrule(lr){5-7}
\rowcolor{tableheader}
\textbf{Method} & \textbf{M} & \textbf{C} & \textbf{Tot./30}
                 & \textbf{M} & \textbf{C} & \textbf{Tot./30}
                 & \textbf{/60} \\
\midrule
DifuzzRTL~\cite{hur2021difuzzrtl}      & 5 & 5 & 10  & 9 & 6 & 15  & 25 \\
ProcessorFuzz~\cite{canakci2023processorfuzz} & 5 & 5 & 10 & 9 & 6 & 15 & 25 \\
Cascade~\cite{solt2024cascade}          & \underline{7} & 4 & \underline{11} & \underline{10} & 5 & 15 & \underline{26} \\
\rowcolor{oursrow}
\textbf{HiFuzz (Ours)}                  & \textbf{7} & \textbf{5} & \textbf{12} & \textbf{12} & \textbf{6} & \textbf{18} & \textbf{30} \\
\midrule
\textit{$\Delta$ vs.\ Cascade}          & $+0$ & $+1$ & $+1$ & $+2$ & $+1$ & $+3$ & $+4$ \\
\bottomrule
\end{tabular}
}
\end{table}

We use the Encarsia benchmark~\cite{bolcskei2025encarsia}, which models CPU bugs as signal or logic-expression mix-ups and broken conditionals, then formally checks that each injected bug can produce an architectural deviation. From its EnCorpus set we run the 30 bugs per DUT (15 Mix-ups + 15 Broken Conditionals; IDs in Table~\ref{tab:bug-mapping} of the Appendix), giving every fuzzer a 24-hour budget per bug on one CPU core.

Table~\ref{tab:bug-detection} summarizes the outcome. On Rocket, all fuzzers remain within a narrow range. On BOOM, HiFuzz detects \textbf{18/30} bugs, while every baseline detects 15. Across both DUTs, HiFuzz detects 30 bugs versus 26 for Cascade.

Encarsia's case studies show that missed bugs often require specific ISA features, CSR side-effect cases, targeted input values, or pressure on microarchitectural structures such as issue queues, buffers, and reorder buffers~\cite{bolcskei2025encarsia}. HiFuzz's advantage on BOOM is consistent with this observation: its hierarchical policy generates broader ISA/CSR scenarios and more structured stress sequences than the baselines under the same time budget.

Beyond Encarsia, we also ran a sanity-check campaign on CVA6 against previously reported real bugs from the literature. HiFuzz triggered the known CVA6 bugs covered by Cascade~\cite{solt2024cascade} and GenHuzz~\cite{wu2025genhuzz}, which provides supporting evidence beyond injected-bug corpora.
\FloatBarrier

\section{Limitations}
HiFuzz is evaluated on open-source RISC-V cores of up to roughly one million gates, and its current implementation targets single-hart program generation without multi-hart synchronization or the A extension, so concurrency-heavy bug classes such as memory-ordering violations remain out of scope. The framework also assumes access to module-level coverage signals and incurs additional encoder/RL overhead relative to purely constructive fuzzers.

Porting HiFuzz to another ISA requires replacing the ISA-specific parts of the Structured Tokenizer and constrained generator: opcode categories, operand roles, register classes, privilege/exception rules, and legal addressing forms. The BB-Sim idea is not tied to RISC-V, but mature ISAs such as ARM or x86 would need richer instruction metadata to handle variable-length encodings, complex addressing modes, implicit operands, flags, and instruction-specific side effects. Porting to a proprietary core also requires a reference model or checker and coverage access at a useful granularity. If only global coverage is available, ACRM can degrade to a global reward, but it loses its ability to redirect effort toward under-covered modules. Finally, because full 24-hour multi-fuzzer campaigns are expensive, the reported coverage curves are not averaged over many random seeds; broader multi-seed reporting with mean and standard deviation is useful future work for quantifying run-to-run variance.

\section{Conclusion}
We presented HiFuzz, a hierarchical RL framework for hardware fuzzing that pairs a macro-level Program Agent with a micro-level Basic Block Agent, driven by a Semantic-Aware BB Encoder for dense intrinsic feedback and an ACRM for balanced module-level coverage. Across three RISC-V cores, HiFuzz consistently outperforms state-of-the-art fuzzers in coverage efficiency and bug detection, with the advantage widening on more complex DUTs.

\bibliographystyle{IEEEtran}
\bibliography{ref}

\appendices
\section{Implementation Configuration Details}\label{sec:impl_appendix}
This appendix complements Section~\ref{sec:impl-details} with the full state/action schemas, network sizes, optimizer settings, and simulation-environment details used in all evaluations.

\textbf{State and Action Spaces.} The Program Agent observes an 11-dimensional state $S_P$ that summarizes recent generation behavior (BB completion statistics, BB quality metrics, and exploration progress) and emits a discrete action $A_P$ with $1{,}375$ combinations formed by the cross-product of $5$ memory-footprint levels, $5$ BB-count levels, $11$ per-BB length distributions, and $5$ instruction-type distributions. The Basic Block Agent observes a 12-dimensional state $S_{BB}$ (remaining instruction budget, current privilege level, and history of recently generated instruction types) and emits a continuous action $A_{BB}$ that parameterizes the instruction-type distribution and BB-termination distribution for the current BB.

\textbf{Networks and Optimization.} The Program Agent is a 3-layer MLP with 128-wide hidden layers trained with Rainbow DQN at learning rate $1\mathrm{e}{-4}$. The Basic Block Agent is an up-to-5-layer MLP with 64-wide hidden layers trained with PPO at learning rate $1\mathrm{e}{-3}$, initial action standard deviation $\texttt{action\_std}{=}0.5$ (annealed to a minimum of $0.1$), intrinsic/extrinsic reward coefficients $\alpha_{\text{int}}{=}1$ and $\alpha_{\text{ext}}{=}2$, and extrinsic-baseline diminishing factor $\gamma{=}0.75$. The Actor and Critic share a common trunk but use separate value heads for intrinsic and extrinsic returns, as described in Section~\ref{subsubsec:bb-agent}.

\textbf{Fuzzing Environment.} HiFuzz integrates with Cocotb for testbench orchestration, Verilator~\cite{verilator2024} for RTL simulation, and Spike as the ISA reference model. Coverage is collected via instrumentation-based primitives at the RTL level---Line, Toggle, MUX~\cite{laeufer2018rfuzz}, and Control Register~\cite{hur2021difuzzrtl}---and aggregated per module for the ACRM. Episode budgets, the ACRM $\beta$ schedule, and the BB Encoder checkpoint paths are all configurable through a single YAML file, which makes every experiment in this paper reproducible by swapping configuration files without code changes.

\textbf{Platform.} All experiments run on an Intel Xeon Gold 6246R server with one NVIDIA A100 GPU. Every fuzzing campaign uses one CPU core per fuzzer to keep simulation cost comparable across baselines.

\section{BB Encoder Model Selection}
\label{sec:encoder_comparison}

We evaluated multiple sequence model architectures for the Basic Block Encoder, including Bi-LSTM, GPT-2, and RWKV-7. Each architecture offers different trade-offs among sequence inductive bias, optimization stability, model size, and inference latency.

\textbf{Bi-LSTM.} Our encoder employs a 2-layer bidirectional LSTM with attention-based pooling. The bidirectional structure captures context from both forward and backward directions, essential for understanding instruction dependencies within a basic block. Architecture details:
\begin{itemize}
    \item Embedding Dimension: 64
    \item LSTM Hidden Dimension: 128 (bidirectional $\rightarrow$ 256 output)
    \item Number of Layers: 2
    \item Pooling Mechanism: Attention-based pooling
    \item Output Dimension: 96 (L2-normalized)
\end{itemize}

\textbf{GPT-2.} We evaluated two variants of the Transformer-based GPT-2 architecture: GPT-2-small (6 layers, 768 hidden) and GPT-2 (12 layers, 1024 hidden). Transformers are expressive sequence models, but their modeling strength is concentrated in longer-range interactions than those typically required by basic blocks, and their larger parameter counts make scratch training under our two-stage curriculum less attractive than a smaller recurrent encoder.

\textbf{RWKV-7.} RWKV is a recent architecture that combines recurrent computation with Transformer-like mixing through linear attention mechanisms. It is a compelling general-purpose sequence model, but in our setting its capacity remains substantially larger than Bi-LSTM, making it a less economical choice for a basic-block encoder trained from scratch.

Table~\ref{tab:encoder_comparison} compares the model sizes and CPU/GPU inference latency of the evaluated encoders.

\begin{table}[!htbp]
\centering
\caption{Comparison of encoder model architectures.}
\label{tab:encoder_comparison}
\setlength{\tabcolsep}{3.5pt}
\renewcommand{\arraystretch}{1.2}
\footnotesize
\begin{adjustbox}{max width=\columnwidth}
\begin{tabular}{lcccc}
\toprule
\rowcolor{tableheader}
\textbf{Model} & \textbf{Params} & \textbf{Size (MB)} & \textbf{CPU Infer.} & \textbf{GPU Infer.} \\
\rowcolor{tableheader}
 &  &  & \textbf{(ms/batch)} & \textbf{(ms/batch)} \\
\midrule
\rowcolor{oursrow}
Bi-LSTM (Ours)  & \textbf{0.69M}  & \textbf{2.65}   & \textbf{52.41} & \textbf{3.71} \\
GPT-2-small     & 7.57M  & 28.86  & 67.83 & \textbf{4.32} \\
GPT-2           & 43.61M & 166.35 & 150.46 & 8.59 \\
RWKV-7          & 90.97M & 347.00 & 1452.04 & 16.34 \\
\bottomrule
\end{tabular}
\end{adjustbox}
\end{table}

\textbf{Selection Rationale.} We selected Bi-LSTM based on two key considerations:
\begin{enumerate}
    \item \textbf{Sequence-Structure Match:} Basic blocks are short, strictly ordered, single-entry single-exit sequences. Their semantics depend primarily on local instruction neighborhoods, operand roles, and forward/backward dependencies, which align naturally with a bidirectional recurrent encoder and reduce the need for heavier long-context attention machinery.
    \item \textbf{Compact and Sufficient:} Our encoder is trained from scratch with a two-stage curriculum rather than initialized from large external corpora. In this regime, Bi-LSTM is easier to optimize and less parameter-hungry than the Transformer and RWKV alternatives, while Fig.~\ref{fig:bbencoder_loss} shows that both training stages converge smoothly in practice. Table~\ref{tab:encoder_comparison} further shows that Bi-LSTM is not only the smallest model, but also the fastest on both CPU and GPU in our setting. Taken together, these results indicate that there is no practical need to move to a substantially larger model for this basic-block sequence task.
\end{enumerate}

\section{Generated Program Quality}\label{sec:prog_quality_appendix}
\begin{figure}[!htbp]
    \centering
    \includegraphics[width=\columnwidth]{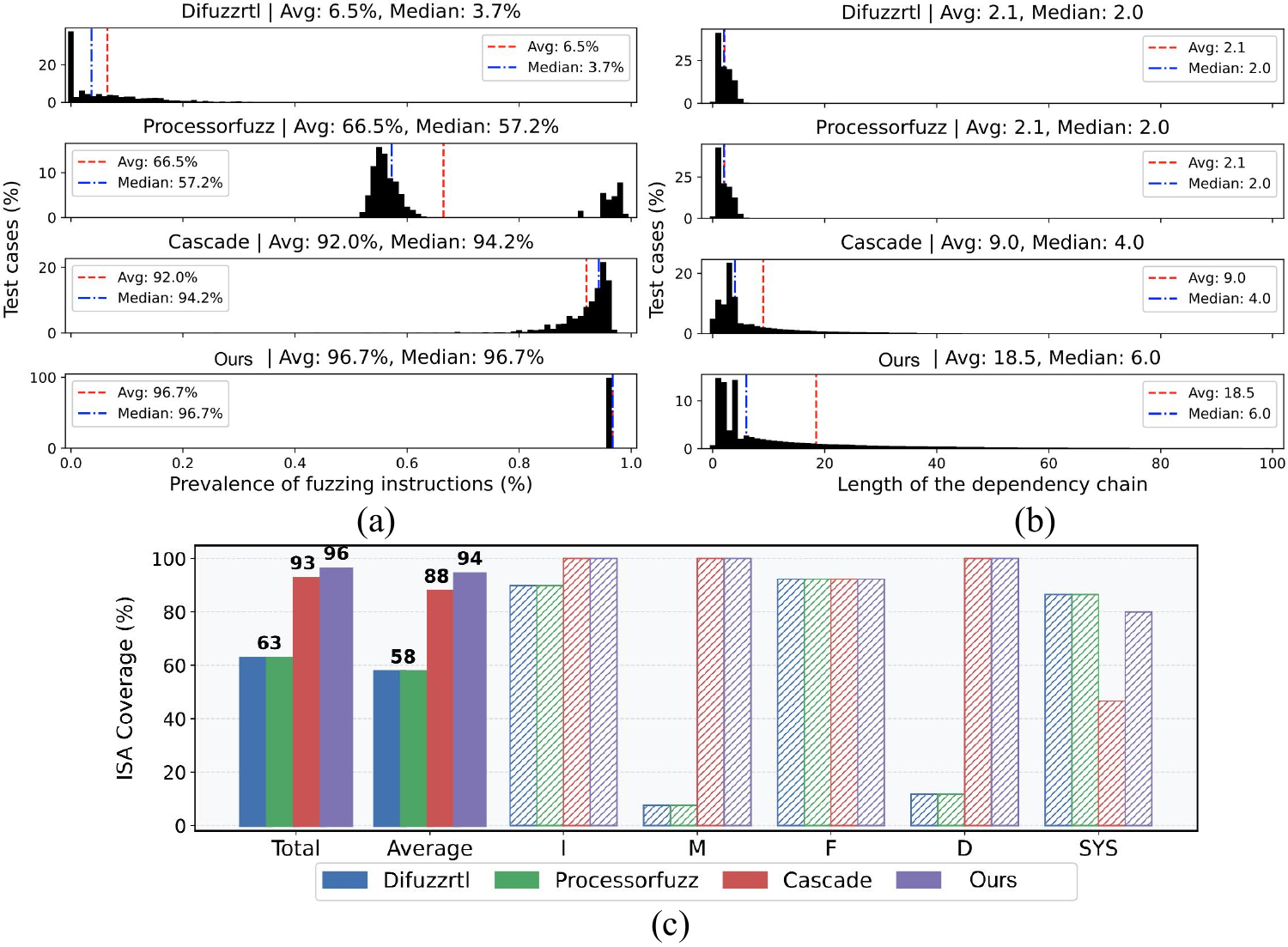}
    \caption{Generated program quality across fuzzers: (a) prevalence, (b) dependency-chain length, and (c) ISA opcode coverage.}
    \label{fig:program_quality}
\end{figure}

To characterize the quality of the programs that each fuzzer actually emits---independently of whether those programs subsequently trigger coverage---we compare DifuzzRTL~\cite{hur2021difuzzrtl}, ProcessorFuzz~\cite{canakci2023processorfuzz}, Cascade~\cite{solt2024cascade}, and HiFuzz over the same $24$-hour campaign along three complementary dimensions: \textit{Prevalence}, \textit{Dependency}, and \textit{ISA Coverage}.

\textit{Prevalence} is the fraction of emitted instructions that actually contribute to fuzzing, as opposed to scaffolding/overhead. As shown in Fig.~\ref{fig:program_quality}(a), DifuzzRTL's stream is dominated by overhead (average $6.5\%$ prevalence), ProcessorFuzz improves to $66.5\%$, Cascade reaches $92.0\%$, and HiFuzz attains near-saturation at both average and median $96.7\%$. This confirms that letting the Program Agent decide program structure, and the Basic Block Agent decide per-BB instruction mix, concentrates generation budget on semantically meaningful instructions rather than on boilerplate.

\textit{Dependency} reports the length of instruction dependency chains, which determines how deeply a test can stress pipelined micro-architectural corner cases. Fig.~\ref{fig:program_quality}(b) shows that DifuzzRTL and ProcessorFuzz stall at an average chain length of $2.1$, Cascade reaches $9.0$, and HiFuzz extends to an average of $18.5$ (median $6.0$). The long tail comes from HRL decisions that deliberately chain operand reuses across BBs, which neither a pure mutator nor a purely constructive generator tends to produce.

\textit{ISA Coverage} measures the functional breadth of the emitted instruction mix over the RISC-V G subset (excluding the A extension, since this paper focuses on single-hart generation). Fig.~\ref{fig:program_quality}(c) shows that DifuzzRTL and ProcessorFuzz cover $63\%$, Cascade reaches $93\%$, and HiFuzz reaches $96\%$, with most of the gap explained by HiFuzz's broader exercising of system-level instructions (CSR operations and privilege-mode transitions).

\section{Bug Detection Details}

Following Encarsia's taxonomy~\cite{bolcskei2025encarsia}, we group injected bugs by \emph{structural root cause} rather than by high-level symptom. \emph{Signal Mix-ups} capture wrong driver, operand, or expression choices in assignments and datapath logic; they often manifest architecturally as corrupted values, wrong control decisions, or misrouted microarchitectural state updates. \emph{Broken Conditionals} capture missing guards, missing cases, or overly permissive predicates; these bugs typically expose exceptional cases, state-dependent checks, or privilege conditions that should have blocked an operation.

Each EnCorpus design instance contains exactly one injected bug, and Encarsia uses formal checks to ensure that this transformation can induce an architecturally observable deviation. Therefore, a successful detection in Tables~\ref{tab:bug-mapping} and~\ref{tab:bug-detection-detailed} should be interpreted as the fuzzer's ability to generate a test that propagates the injected RTL fault to ISA-visible behavior, not merely to toggle the local logic containing the bug.

\begin{table*}[!t]
\centering
\caption{Bug mapping for different designs and bug types.}
\label{tab:bug-mapping}
\renewcommand{\arraystretch}{1.2}
\setlength{\tabcolsep}{2.2pt}
\normalsize
\begin{adjustbox}{max width=\textwidth}
\begin{tabular}{clccccccccccccccc}
\toprule
\rowcolor{tableheader}
\multirow{2}{*}{\textbf{Design}} & \multirow{2}{*}{\textbf{Bug Type}} & \multicolumn{15}{c}{\textbf{Bug IDs}} \\
\cmidrule(lr){3-17}
\rowcolor{tableheader}
 & & \textbf{1} & \textbf{2} & \textbf{3} & \textbf{4} & \textbf{5} & \textbf{6} & \textbf{7} & \textbf{8} & \textbf{9} & \textbf{10} & \textbf{11} & \textbf{12} & \textbf{13} & \textbf{14} & \textbf{15} \\
\midrule
\multirow{2}{*}{\textbf{Rocket}} & Mix-ups & 144 & 293 & 347 & 361 & 640 & 673 & 767 & 774 & 804 & 826 & 832 & 858 & 863 & 1081 & 1094 \\
 & Conditionals & 14 & 28 & 46 & 66 & 71 & 277 & 285 & 295 & 305 & 394 & 397 & 408 & 588 & 633 & 718 \\
\midrule
\multirow{2}{*}{\textbf{BOOM}} & Mix-ups & 27 & 56 & 137 & 148 & 228 & 589 & 604 & 657 & 741 & 787 & 954 & 990 & 1028 & 1073 & 1236 \\
 & Conditionals & 3 & 6 & 137 & 160 & 168 & 176 & 187 & 195 & 200 & 513 & 517 & 530 & 708 & 727 & 975 \\
\bottomrule
\end{tabular}
\end{adjustbox}
\end{table*}

\begin{table*}[!t]
\centering
\caption{Detailed bug detection results. \ding{51}: detected, --: not detected.}
\label{tab:bug-detection-detailed}
\setlength{\tabcolsep}{1.6pt}
\renewcommand{\arraystretch}{1.02}
\normalsize
\begin{adjustbox}{max width=\textwidth}
\begin{tabular}{c|cccc|cccc|cccc|cccc}
\toprule
\rowcolor{tableheader}
\multirow{3}{*}{\textbf{\#}} & \multicolumn{4}{c}{\textbf{Rocket Mix-ups}} & \multicolumn{4}{c}{\textbf{Rocket Cond.}} & \multicolumn{4}{c}{\textbf{BOOM Mix-ups}} & \multicolumn{4}{c}{\textbf{BOOM Cond.}} \\
\cmidrule(lr){2-5} \cmidrule(lr){6-9} \cmidrule(lr){10-13} \cmidrule(lr){14-17}
\rowcolor{tableheader}
 & DF & PF & CA & HF & DF & PF & CA & HF & DF & PF & CA & HF & DF & PF & CA & HF \\
\midrule
1  & -- & -- & \ding{51} & \ding{51} & -- & -- & -- & -- & \ding{51} & \ding{51} & \ding{51} & \ding{51} & \ding{51} & \ding{51} & \ding{51} & \ding{51} \\
2  & \ding{51} & \ding{51} & \ding{51} & \ding{51} & -- & -- & \ding{51} & \ding{51} & \ding{51} & \ding{51} & \ding{51} & \ding{51} & \ding{51} & \ding{51} & \ding{51} & \ding{51} \\
3  & \ding{51} & \ding{51} & \ding{51} & \ding{51} & -- & -- & -- & -- & -- & -- & -- & -- & -- & -- & -- & -- \\
4  & \ding{51} & \ding{51} & \ding{51} & \ding{51} & -- & -- & -- & -- & \ding{51} & \ding{51} & -- & \ding{51} & -- & -- & -- & -- \\
5  & -- & -- & -- & -- & -- & -- & -- & -- & \ding{51} & \ding{51} & -- & \ding{51} & -- & -- & -- & -- \\
6  & -- & -- & \ding{51} & \ding{51} & \ding{51} & \ding{51} & \ding{51} & \ding{51} & -- & -- & \ding{51} & \ding{51} & -- & -- & -- & -- \\
7  & -- & -- & -- & -- & \ding{51} & \ding{51} & -- & -- & -- & -- & \ding{51} & \ding{51} & \ding{51} & \ding{51} & \ding{51} & \ding{51} \\
8  & -- & -- & -- & -- & -- & -- & -- & -- & \ding{51} & \ding{51} & \ding{51} & \ding{51} & -- & -- & -- & -- \\
9  & -- & -- & -- & -- & -- & -- & -- & -- & -- & -- & \ding{51} & \ding{51} & -- & -- & -- & -- \\
10 & -- & -- & -- & -- & \ding{51} & \ding{51} & -- & \ding{51} & \ding{51} & \ding{51} & \ding{51} & \ding{51} & -- & -- & -- & -- \\
11 & -- & -- & -- & -- & -- & -- & -- & -- & \ding{51} & \ding{51} & \ding{51} & \ding{51} & \ding{51} & \ding{51} & \ding{51} & \ding{51} \\
12 & -- & -- & -- & -- & \ding{51} & \ding{51} & \ding{51} & \ding{51} & \ding{51} & \ding{51} & -- & -- & \ding{51} & \ding{51} & \ding{51} & \ding{51} \\
13 & \ding{51} & \ding{51} & \ding{51} & \ding{51} & -- & -- & -- & -- & -- & -- & -- & -- & -- & -- & -- & -- \\
14 & -- & -- & -- & -- & -- & -- & -- & -- & -- & -- & \ding{51} & \ding{51} & -- & -- & -- & -- \\
15 & \ding{51} & \ding{51} & \ding{51} & \ding{51} & \ding{51} & \ding{51} & \ding{51} & \ding{51} & \ding{51} & \ding{51} & \ding{51} & \ding{51} & \ding{51} & \ding{51} & -- & \ding{51} \\
\bottomrule
\end{tabular}
\end{adjustbox}
\end{table*}

\end{document}